\documentclass[aps,prr,twocolumn,superscriptaddress,longbibliography]{revtex4-2}

\usepackage[utf8]{inputenc}

\usepackage{mathtools}
\usepackage{amsfonts}
\usepackage{mathrsfs}
\usepackage{physics}
\usepackage{slashed}
\usepackage{tensor}
\usepackage{dsfont}
\usepackage{bbm}

\usepackage{graphicx}
\usepackage{float}
\usepackage{array}
\usepackage[abs]{overpic}

\usepackage{placeins}

\usepackage{makecell}

\usepackage{xspace}
\usepackage{siunitx}
\usepackage{xfrac}
\usepackage{hyperref}
\usepackage[nameinlink]{cleveref}

\usepackage{xifthen}
\usepackage{xcolor}
\hypersetup{
	colorlinks,
	linkcolor={red!75!black},
	citecolor={blue!75!black},
	urlcolor={blue!75!black}
}

\usepackage{booktabs}

\newcolumntype{C}{>{$}c<{$}}
\AtBeginDocument{
	\heavyrulewidth=.08em
	\lightrulewidth=.05em
	\cmidrulewidth=.03em
	\belowrulesep=.65ex
	\belowbottomsep=0pt
	\aboverulesep=.4ex
	\abovetopsep=0pt
	\cmidrulesep=\doublerulesep
	\cmidrulekern=.5em
	\defaultaddspace=.5em
}

\newcommand{\gettitle}{Critical aging and relaxation dynamics in long-range systems}

\newcommand{\getHeidelbergAffiliation}{\affiliation{Institut f{\"u}r Theoretische Physik, Universit{\"a}t Heidelberg, Philosophenweg 16, 69120 Heidelberg, Germany}}
\newcommand{\getZuerichAffiliation}{\affiliation{Institut f{\"u}r Theoretische Physik, ETH Z{\"u}rich, Wolfgang-Pauli-Str. 27, 8093 Z{\"u}rich, Switzerland}}

\hypersetup{
	bookmarksopen=true,
	bookmarksopenlevel=2,
	bookmarksnumbered=true
}

\begin{document}
	
\title{\gettitle}

\author{Valerio Pagni}\getZuerichAffiliation
\thanks{Contact author: vpagni@ethz.ch}
\author{Friederike Ihssen}\getHeidelbergAffiliation
\author{Nicolò Defenu}\getZuerichAffiliation

\begin{abstract}
    We study the dynamical scaling of long-range $\mathrm{O}(N)$ models after a sudden quench to the critical temperature, using the functional renormalization group approach. We characterize both short-time aging and long-time relaxation as a function of the symmetry index $N$, the interaction range decay exponent $\sigma$ and the dimension $d$. Our results substantially improve on perturbative predictions, as demonstrated by benchmarks against Monte Carlo simulations and the large-$N$ limit. Finally, we demonstrate that long-range systems increase the performance of critical heat engines with respect to a local active medium.
\end{abstract}

\maketitle
 
\section{Introduction}\label{sec:intro}

The study of long-range interacting systems has recently garnered renewed interest due to their relevance in non-equilibrium statistical mechanics~\cite{campa2009statistical} and quantum many-body physics~\cite{defenu2023long,defenu2024out}, as well as their experimental realizations in Rydberg atoms, trapped-ion systems, and cold atoms in cavities~\cite{browaeys2020many,britton2012engineered,lahaye2009physics,ritsch2013cold,defenu2023long}. In this work, we focus specifically on the critical dynamics of long-range $N$-vector spin models -- systems in which two-body ferromagnetic interactions decay as a power law, $r^{-(d+\sigma)}$, with the distance $r$ between two classical spins. Here, $d$ denotes the spatial dimension, while $\sigma$ controls the interaction decay rate.

Consider the dynamics without conservation laws~\cite{hohenberg1977theory} of such a spin system that follows a sudden quench of the temperature from the deeply disordered phase to the critical point -- for quenches below the critical temperature, see e.g.~\cite{christiansen2020aging}. 
While the system undergoes the standard critical relaxation towards equilibrium according to the dynamical exponent $z$, cf.~\cite{hohenberg1977theory}, in the long-time limit, the dynamics of correlation functions at shorter times exhibits distinctive non-equilibrium features~\cite{janssen1989new,janssen1992renormalized,chen2000short,calabrese2005ageing,ihssen2025nonperturbative}, which we refer to as \textit{critical aging}. In general, aging behavior is a hallmark of systems with slow, non-equilibrium dynamics, where physical properties evolve in a history-dependent manner. In contrast to systems featuring a quick relaxation to equilibrium, aging systems display a non-trivial evolution of their correlation and response functions, characterized by a lack of time-translation invariance~\cite{henkel2007ageing,henkel2011non}. This phenomenon is observed in a variety of systems, including glassy materials and spin systems~\cite{cugliandolo2002dynamics}. 

In the context of ferromagnetic spin models, aging in the short-time dynamics is characterized by a non-equilibrium critical exponent $\theta$, which has been extensively studied for short-range interacting systems~\cite{calabrese2005ageing}. On the other hand, considering their much less investigated long-range counterparts provides two key advantages: control over the critical exponents, and a tunable time window during which the characteristic aging properties of the dynamics can be observed. Specifically, the aging exponent $\theta$ depends continuously on the decay parameter $\sigma$, which interpolates between mean-field (small $\sigma$) and short-range (large $\sigma$) behavior~\cite{defenu2023long}. Likewise, the crossover time $t_{\text{cross}}$, marking the transition from short-time behavior to long-time relaxation, can be adjusted accordingly.

Our contribution is threefold:
(i) By means of the non-perturbative renormalization group (RG), we characterize the dynamical universality in the whole range of the power-law parameter $\sigma \in (0,\infty)$, overcoming several limitations of the perturbative RG approach of~\cite{chen2000short}. The calculation of the critical exponents $z$ and $\theta$ is carried out for several values of $N$, the number of components of the spins. The Monte Carlo (MC) simulations of the one-dimensional long-range Ising model~\cite{uzelac2008short,rodriguez2011study} represent a benchmark for the accuracy of our results for $N=1$.
(ii) The existence of an effective fractional dimension $D$ that enables the reconstruction of the \textit{equilibrium} critical properties of the long-range models from a corresponding local model in dimension $D$, has been explored in~\cite{banos2012correspondence,angelini2014relations,defenu2015fixed,defenu2017criticality,solfanelli2024universality}. In the present work, we extend this framework to \textit{out-of-equilibrium} critical behavior, by providing a dictionary between short- and long-range values of the exponents $z$ and $\theta$. 
(iii) Following the suggestion of~\cite{campisi2016power} of using critical systems as the working medium of a thermodynamic heat engine, we extend this proposal to long-range systems. In this case, through the calculation of what we dub `performance-rate exponent' $\pi_{\rm th} = \alpha-z\nu$, where $\alpha$ and $\nu$ are equilibrium critical exponents, we observe a thermodynamic advantage in the scaling of the performance rate, defined below, over the case with nearest-neighbor interactions.

The paper is structured as follows: In~\Cref{sec:model} we introduce the model and the temperature quench, including a discussion of the exactly solvable large-$N$ limit. In~\Cref{sec:effdim} we present the effective dimension approach and its proposed extension to non-equilibrium scenarios. In~\Cref{sec:frg} we describe the functional renormalization group (fRG) approach that enables us to calculate the dynamical exponent $z$ and the aging exponent $\theta$. Our main results -- including those of the performance rate scaling -- are presented in~\Cref{sec:results}, followed by the conclusion and outlook in~\Cref{sec:conclusions}.

\section{Temperature quench of the long-range $\mathrm{O}(N)$ models}\label{sec:model}

We consider a long-range lattice model given by the $\mathrm{O}(N)$-symmetric classical Hamiltonian~\cite{defenu2023long}
\begin{equation}\label{LR Hamiltonian}
	H = -\frac{1}{2} \sum_{i, j} J_{ij} \boldsymbol{S}_i \cdot \boldsymbol{S}_j \,,
	\qquad J_{ij} \propto \frac{1}{|i-j|^{d+\sigma}}\,,
\end{equation}
where the non-negative matrix elements $J_{ij}$ decay algebraically with the distance between the sites $i$ and $j$, $d$ is the spatial dimension, $\sigma>0$ parametrizes the range of the interactions, and $\boldsymbol{S}_i$ are $N$-component vectors of unit length.

The system is initially prepared in a high-temperature disordered configuration with a non-vanishing but small magnetization $M_0$. This state is well described by mean-field theory, as critical fluctuations are absent. Then, at time $t=t_0$, the system is suddenly brought to its critical temperature $T_c$, in the absence of any external magnetic fields.
As in the short-range case~\cite{janssen1989new,janssen1992renormalized,calabrese2005ageing}, we expect to observe some dynamics that depend on microscopic details of the system and of the initial state very shortly after the quench, until a universal behavior emerges at intermediate -- but macroscopically \textit{short} -- times, see also~\cite{chen2000short}. The latter short-time universality manifests itself as critical aging in two-point functions, e.g.~the response function:
\begin{equation}\label{two-point function aging}G^R(\boldsymbol{q},t,t') \sim (t/t')^{\theta} f_R(q(t-t')^{1/z}, t'/t)\,,
	\quad t' \to t_0\,,
\end{equation}
where $f_R$ is a scaling function and $\theta$ is the aging exponent~\cite{calabrese2005ageing,ihssen2025nonperturbative}, which is independent of the equilibrium critical exponent for the dynamics we are interested in (i.e.~model A, as explained later). Alternatively, one can look at the initial buildup in the magnetization $M(t)$, intuitively explained~\cite{janssen1992renormalized} by noticing that the target temperature of the quench, $T_c$, is lower than the mean-field critical temperature $T_c^{\text{mf}}$, and therefore there is an initial ordering of the spins until a time $\sim t_{\text{cross}}$ at which critical correlations have been established. The behavior of magnetization is
\begin{equation}\label{magnetization initial slip}
	M(t) = M_0 t^{\theta'} f_M(M_0 t^{\theta'+\beta/(\nu z)})\,,
\end{equation}
where $\theta' = \theta + (\sigma-z)/z$ -- this relation being the generalization of the short-range one~\cite{janssen1989new} -- is called initial slip exponent, and the corresponding scaling function behaves as $f_M(0) = 1$ and $f_M(u \to \infty) \sim u^{-1}$. The exponents $\beta$ and $\nu$ are related to the magnetization and correlation length $\xi$ at equilibrium. As a consequence of the scaling \eqref{magnetization initial slip}, we can estimate the crossover time to be given by $t_{\text{cross}} \propto M_0^{-\psi}$, where $\psi = [\theta'+\beta/(\nu z)]^{-1}$. In the long-time limit the evolution of the system eventually crosses over to the close-to-equilibrium critical dynamics characterized by the divergence of the relaxation time $\tau_{\text{relax}} \sim \xi^z$, where $z$ is the dynamical exponent.

In the limit $\sigma \to \infty$ of the Hamiltonian~\eqref{LR Hamiltonian} we retrieve the usual nearest-neighbor short-range $\mathrm{O}(N)$ vector model. In fact, the same \textit{equilibrium} critical behavior as the short-range model is already recovered for $\sigma \geq \sigma_* = 2-\eta_{\text{SR}}$ where $\eta_{\text{SR}}$ is the anomalous dimension of the corresponding short-range model. On the other hand, when $0<\sigma\leq \sigma_{\text{mf}}=d/2$ the range of interactions is so large that mean-field critical behavior is recovered. The remaining interval $\sigma \in (\sigma_{\text{mf}},\sigma_*)$ exhibits genuine long-range critical behavior, where the critical exponents are a continuous function of the parameter $\sigma$~\cite{defenu2023long}. The universal properties of the Hamiltonian~\eqref{LR Hamiltonian} correspond to those of the effective field theory
\begin{align}\label{phi4 action}
	\mathcal{H}[\boldsymbol{\varphi}] = \int_{\boldsymbol{x} } \left\{ \frac12 (\nabla^{\frac{\sigma}{2}} \boldsymbol{\varphi})^2 + \frac{\tau}{2} \boldsymbol{\varphi}^2 + \frac{g}{4!}(\boldsymbol{\varphi}^2)^2 \right\}\,,
\end{align}
where $\int_{\boldsymbol{x}} \equiv \int d^d \boldsymbol x$ and $\boldsymbol{\varphi} = (\varphi_1, \dots, \varphi_N)$ is a continuous field. In the infrared, the fractional gradient can be interpreted in momentum space as follows: $\int_{\boldsymbol{x}} (\nabla^{\frac{\sigma}{2}} \boldsymbol{\varphi}(\boldsymbol{x}))^2 = \int_{\boldsymbol{q}} q^\sigma \boldsymbol{\varphi}(-\boldsymbol{q}) \cdot \boldsymbol{\varphi}(\boldsymbol{q})$, where $\int_{\boldsymbol{q}} \equiv (2\pi)^{-d} \int d^d q$ and $\boldsymbol{\varphi}(\boldsymbol{q})$ is the Fourier transform of $\boldsymbol{\varphi}(\boldsymbol{x})$. At tree level the relevance of long-range couplings in the low-energy limit is obtained by comparing the scaling dimension of the fractional gradient $k^{\sigma}$ with the one of the conventional local gradient term $k^{2}$, where $k$ is the infrared cutoff scale. The boundary $\sigma_{*}=2-\eta_{\rm SR}$ is obtained by comparing the scaling dimension of the long-range operator with the one of the renormalized local kinetic term $k^{2-\eta_{\rm SR}}$, as first argued by J. Sak~\cite{sak1973recursion}.

Sak's crossover scenario has been thoroughly investigated over the years, with many studies focusing on the Ising model ($N=1$): Refs.~\cite{picco2012critical,blanchard2013influence} predict the crossover to remain at $\sigma^* = 2$ even for the full theory, but this finding has been attributed to the difficulty of capturing logarithmic corrections close to the boundary~\cite{angelini2014relations}. Sak’s picture is confirmed by MC simulations in two dimensions (2D)~\cite{luijten2002boundary, angelini2014relations} and RG approaches\,\cite{brezin2014crossover, defenu2015fixed}. Conformal perturbation theory studies agree with Sak's criterion in both two and three dimensions~\cite{behan2017long, behan2017scaling}. More recently, interest has been shifted to $d=1$, where the crossover at $\sigma^{*}=1$ has been investigated by both conformal perturbation theory~\cite{benedetti2025one} and functional RG~\cite{pagni2025one}.
 
In the same spirit, we set out to study the critical dynamics of the system from a field-theoretical perspective. In particular, we investigate the above scenario applied to critical behavior of long-range systems away from thermal equilibrium. We remark that the dynamics we are considering is not directly given by the Hamiltonian~\eqref{LR Hamiltonian}. Rather, it is implemented phenomenologically at the mesoscopic level by requiring that the system relaxes to a Gibbs distribution and that possible conservation laws are retained by constraining the ensemble~\cite{hohenberg1977theory,cardy1996scaling}. We work in the absence of any conservation laws, i.e.~we take into account the dynamics of model A in the traditional classification~\cite{hohenberg1977theory}. Therefore, one writes a Langevin equation of the form
\begin{align}\label{Langevin}
	\partial_t {\varphi}(t,\boldsymbol{x})= - \mathcal{D} \frac{\delta \mathcal{H}[{\varphi}]}{\delta {\varphi}(t,\boldsymbol{x})} + \zeta(t,\boldsymbol{x})\,,
\end{align}
where $\mathcal{D}$ is a constant relaxation rate and $\zeta$ is a zero-mean Markovian and Gaussian noise with correlation
\begin{align}
	\langle \zeta(t_1,\boldsymbol x_1)\zeta(t_2,\boldsymbol x_2) \rangle = 2 \Omega \, \delta(\boldsymbol x_1-\boldsymbol x_2) \delta(t_1-t_2)\,.
\end{align}
The Einstein relation \cite{tauber2014critical} is realized when the amplitude of the noise is $\Omega = \mathcal{D}$ (working in units where $k_B T = 1$). Instead of computing observables such as magnetization $\boldsymbol{M}(t) = \langle \boldsymbol{\varphi}(t) \rangle$ by averaging over solutions to the stochastic equation~\eqref{Langevin}, one can recast the problem in terms of functional integrals, by means of the MSRJD~\cite{martin1973statistical, janssen1976on, de1978dynamics} or response field formalism, where auxiliary (response) fields $\tilde{\boldsymbol{\varphi}}$ are introduced for each component of the order-parameter field $\boldsymbol{\varphi}$. Importantly, information about the initial state has to be taken into account in our analysis. This procedure leads to a field-theoretical problem described by the action~\cite{janssen1989new,janssen1992renormalized,calabrese2005ageing}
\begin{align}\label{msr action}
	& S[\varphi,\tilde{\varphi}] = \int_{\boldsymbol{x}} \frac{\tau_0}{2} \left[\boldsymbol{\varphi}_0(\boldsymbol{x})-\boldsymbol{h}(\boldsymbol{x})\right]^2
	\nonumber\\
	& + \int_{t > t_0}  \int_{\boldsymbol{x}} \left\{\tilde{\varphi}_i \partial_{t} \varphi_i + \mathcal{D} \tilde{\varphi}_i  \frac{\delta \mathcal{H}[\boldsymbol{\varphi}]}{\delta \varphi_i} - \tilde{\varphi}_i \Omega \tilde{\varphi}_i \right\}\,,
\end{align}
where summation over repeated indices is implied, $\int_{t > t_0} \equiv \int_{t_0}^{\infty} dt$, and the first line encodes the Gaussian probability distribution of the initial high-temperature state $\boldsymbol{\varphi}_0(\boldsymbol{x}) \equiv \boldsymbol{\varphi}(t_0,\boldsymbol{x})$ with
\begin{align}\label{eq:correlators_initial_time}
	&\langle \boldsymbol{\varphi}_0(\boldsymbol{x})\rangle = \boldsymbol{h}(\boldsymbol{x})\,,
	\nonumber\\
	&\langle \left[\boldsymbol{\varphi}_0(\boldsymbol{x})-\boldsymbol{h}(\boldsymbol{x})\right] \left[\boldsymbol{\varphi}_0(\boldsymbol{x}')-\boldsymbol{h}(\boldsymbol{x}')\right] \rangle = \tau_{0}^{-1} \delta(\boldsymbol{x}-\boldsymbol{x}')\,.
\end{align}
Given that the field $\varphi$ -- as well as $\varphi_0$ -- has mass dimension $(d-\sigma)/2$, one finds that the dimension of $\tau_0$ is $\sigma>0$. Hence, under renormalization $(\tau_{0}^{-1})^* = 0$ is the only fixed point value of the initial correlation length compatible with the normalization of the probability distribution.

Thus the scaling forms~\eqref{two-point function aging} and~\eqref{magnetization initial slip} can be obtained by a RG analysis similar to those of Refs.~\cite{janssen1989new,janssen1992renormalized,calabrese2005ageing}. In particular, from a field-theoretical point of view a new exponent $\theta$ (or $\theta'$) arises due to the fact that the response field at the initial time surface, $\tilde{\boldsymbol{\varphi}}_0 = \tilde{\boldsymbol{\varphi}}(t=t_0)$, has to be renormalized independently of the `bulk' fields and therefore it acquires an anomalous dimension $\tilde{\eta}_0$, see.~\Cref{sec:frg}.

Before considering the strongly interacting critical model, it is convenient to inspect the quadratic model obtained by setting $g = 0$ in~\eqref{phi4 action}. In this case, the correlation and response functions -- defined as $G^C_{0,ij}(\boldsymbol{q},t,t')\equiv\langle\varphi_i(\boldsymbol{q},t)\varphi_j(-\boldsymbol{q},t')\rangle_0=\delta_{ij}G^C_0(\boldsymbol{q},t,t')$ and $G^R_{0,ij}(\boldsymbol{q},t,t')\equiv\langle\varphi_i(\boldsymbol{q},t)\tilde\varphi_j(-\boldsymbol{q},t')\rangle_0=\delta_{ij}G^R_0(\boldsymbol{q},t,t')$ (see e.g.~\cite{tauber2014critical}) -- read
\begin{subequations}\label{gaussian correlators}
    \begin{align}
		G^C_0(\boldsymbol{q},t,t') &= \frac{1}{\omega_q} \left[ e^{-\omega_q |t-t'|} + \left(\frac{\omega_q}{\tau_0}-1\right) e^{-\omega_q (t+t'-2t_0)} \right]
		\label{gaussian correlators C}
		\\
        G^R_0(\boldsymbol{q},t,t') &= \vartheta(t-t') e^{-\omega_q (t-t')}\,,
		\label{gaussian correlators G}
    \end{align}
\end{subequations}
where $\vartheta(\cdot)$ is the Heaviside step function, the dispersion relation $\omega_q$ (with $q=|\boldsymbol{q}|$) is 
\begin{equation}\label{renormalized dispersion relation}
    \omega_q = q^{\sigma} + \tau\,,
\end{equation}
and we have set $\Omega = \mathcal{D}=1$. Comparing~\eqref{two-point function aging} with~\eqref{gaussian correlators G} we obtain $\theta_{\text{mf}} = 0$ as the mean-field value of the aging exponent. Similarly, $z_{\text{mf}} = \sigma$, so that $\theta'_{\text{mf}} = 0$ as well. Hence, it is apparent that critical aging is a genuinely collective phenomenon that requires a non-mean-field description.

A first step towards that is provided by the large-$N$ limit, where the number of field components is taken to be infinite. In fact, following~\cite{janssen1989new,halimeh2021quantum}, we can decouple self-consistently the nonlinearity in the action~\eqref{phi4 action}. This is exact as $N\to\infty$, in which case fluctuations of the variable $N^{-1}  \boldsymbol{\varphi}^2 = N^{-1} \sum_i \varphi_i^2$ are strongly suppressed, see e.g.~\cite{moshe2003quantum}. More precisely, after a rescaling of $g$ by a factor of $N$, we make the replacement
\begin{equation}\label{eq:large N decoupling}
	\frac{g}{N} \sum_{i,j=1}^N \tilde{\varphi}_i \varphi_i \varphi_j \varphi_j \to g \, C(t) \sum_{i=1}^N \tilde{\varphi}_i \varphi_i\,,
\end{equation}
where
\begin{align}
	&C(t) \equiv G^C(\boldsymbol{0},t,t) \notag \\[1ex]
    &\hspace{4mm}= \frac{1}{N} \sum_{i=1}^N \langle \varphi_i(\boldsymbol{x}, t) \varphi_i(\boldsymbol{x}, t) \rangle = \int_{\boldsymbol{q}} C(\boldsymbol{q},t)\,,
\end{align}
is the equal-time correlator. The resulting action is quadratic in the fields, where, however, the quadratic coupling has become time-dependent:
\begin{equation}\label{time-dependent coupling}
	\tau \to \tau + \frac{g}{6} C(t) \equiv \tau_C(t)\,.
\end{equation}
It turns out, as detailed in~\Cref{app:largeN}, that the scaling behavior of the response function is
\begin{equation}\label{critical scaling response function aging}
	G^R(\boldsymbol{q},t,t') = \vartheta(t-t') (t/t')^{\theta} e^{-q^\sigma (t-t')}\,,
\end{equation}
with an aging exponent
\begin{equation}\label{aging exponent largeN} 
	\theta = 1 - \frac{d}{2\sigma}\,,
\end{equation}
which represents an important benchmark for~\Cref{sec:results} in order to test the fRG for $O(N)$ models with $N \gg 1$. Moreover, comparing~\eqref{critical scaling response function aging} with~\eqref{two-point function aging}, we also note that $z = \sigma$ in the large-$N$ limit. As announced, this result already shows that $\theta$ and $z$ can be tuned by changing the range $\sigma$ of the interactions. Notice that~\eqref{aging exponent largeN} can be compared with the large-$N$ estimate of the short-range counterpart of the aging exponent, that reads $\theta_{\text{SR}} = 1-D/4$ for a $D$-dimensional short-range model~\cite{janssen1989new}. The exponents match exactly if $D=2d/\sigma$, where $d$ is the dimension of the \textit{long}-range system. In the following, we explore this equivalence in greater detail and beyond the large-$N$ limit.

\section{Effective dimension approach}\label{sec:effdim}

In thermal equilibrium, the effective dimension approach~\cite{banos2012correspondence,angelini2014relations,defenu2015fixed,defenu2017criticality,solfanelli2024universality} enables us to obtain a pretty accurate (yet not exact) idea of the critical behavior of a long-range model in dimension $d$ with decay exponent $\sigma$ by looking at the corresponding results obtained for a short-range system in the fractional dimension $D = D(\sigma)$. In this section we briefly review this dimensional correspondence and extend it to out-of-equilibrium systems.

Rather than looking at the free energy density as in~\cite{banos2012correspondence,angelini2014relations,solfanelli2024universality}, we prefer to work directly with correlation functions, in order to more clearly reveal the connection with dynamics, where the generating functional obtained from~\eqref{msr action} in the MSRJD framework has a less transparent physical meaning than the equilibrium partition function.

It is well-known from the usual real-space RG applied to finite-size systems, see e.g.~\cite{cardy1996scaling, cardy2012finite}, that the two-point response function for a nearly critical macroscopic system with $\mathcal{N} = L^d$ spins in the bulk obeys
\begin{equation}\label{finite size RG response}
	G(r,t;\{u_\alpha\}) = L^{2(y_h - d)} G(L^{-1} r, L^{-z} t; \{u_\alpha'\})\,,
\end{equation}
after a suitable number of RG iterations. Here $\{u_\alpha\}$ indicate the couplings of the model, among which the relevant ones are the reduced temperature $\tau$ and magnetic field $h$, while $u_\alpha' = L^{y_\alpha} u_\alpha$ are the couplings rescaled according to their RG eigenvalues $y_\alpha$. The relevant couplings have positive eigenvalues $y_\tau$ and $y_h$, related to the critical exponents $\nu$ and $\eta$ by $\nu = 1/y_\tau$ and $\eta = d+2-2y_h$. The dynamical exponent $z$ quantifies the anisotropy between spatial and temporal directions.

The idea of the effective dimension approach is to compare the scaling of the response function between a long-range and a short-range model with the same number of spins $\mathcal{N} = L^d = L_{\text{SR}}^D$. Hereafter, the quantities of the short-range model will be denoted by the label `SR', except $d_{\text{SR}} \equiv D$. Thus, equating the r.h.s.~of~\eqref{finite size RG response} at criticality for the long-range and short-range model yields
\begin{align}
	\frac{2 - \eta(\sigma)}{d} &= \, \frac{2-\eta_{\text{SR}}(D)}{D}\,,
	\label{effective dimension eta}
	\\
	 \frac{z(\sigma)}{d} &= \, \frac{z_{\text{SR}}(D)}{D}
	\label{effective dimension z}\,.
\end{align}
Before discussing how to obtain the effective dimension $D = D(\sigma)$ of the short-range system, we notice that~\eqref{effective dimension z} is already a result pertaining dynamics, even though limited to relaxation close to equilibrium at long times.

In fact, to obtain the effective dimension one needs the additional information that in the long-range case interactions cannot renormalize the non-analytic kinetic term, thus $\eta(\sigma) = 2 - \sigma$ even beyond mean field theory~\cite{sak1973recursion,honkonen1989crossover,honkonen1990critical}. Then, using~\eqref{effective dimension eta}, we obtain that the effective dimension $D = D(\sigma)$ of the SR system is
\begin{equation}\label{effective dimension}
    D = \frac{2-\eta_{\text{SR}}(D)}{\sigma}d\,.
\end{equation}
This is an implicit equation, because one needs to know the value of the critical exponent $\eta_{\text{SR}}$ in fractional dimensions (see e.g.~\cite{codello2013n}) to calculate $D$, as discussed further in~\Cref{sec:results}.

Finally, replacing the expression \eqref{finite size RG response} with the analogous non-equilibrium form of the correlation and response functions given in~\cite{janssen1989new,calabrese2005ageing} (see also~\eqref{critical scaling response function aging}), we are able to confirm the effective dimension relation between long- and short-range models for the aging exponent. In particular, considering the two-time dependence $(t/t')^{\theta}$,
\begin{equation}\label{effective dimension theta}
	\theta(\sigma) = \theta_{\text{SR}}(D(\sigma))\,.
\end{equation}
This generalizes the discussion in the last paragraph of~\Cref{sec:model} to the case where $N$ is finite and, accordingly, the anomalous dimension $\eta_{\text{SR}}(D)$ appearing in~\eqref{effective dimension} is non-vanishing. 

We stress again that, as discussed in
Refs.~\cite{banos2012correspondence,angelini2014relations,defenu2015fixed,defenu2017criticality,behan2017scaling, solfanelli2024universality}, the correspondence between short- and long-range model via the effective dimension is not exact. However, for the two-dimensional long-range Ising model, Ref.~\cite{solfanelli2024universality} tested the effective-dimension prediction for the correlation-length exponent against conformal-bootstrap data and found agreement better than 97\% over the whole parameter range. In the following, we investigate its extension to out-of-equilibrium criticality in two complementary ways. First, in~\Cref{sec:results} we compare long-range and short-range calculations performed independently within the same fRG truncation, in order to test whether the effective dimension correspondence is preserved by our approximation. Second, in~\Cref{subsec:1DLRI} we apply the correspondence to Monte Carlo estimates for short-range models and compare the resulting predictions with long-range Monte Carlo data, thereby providing a benchmark that does not rely on field-theoretical descriptions.

\section{Renormalization group}\label{sec:frg}

In order to obtain the dynamical scaling properties of the long-range model captured by the action~\eqref{msr action}, we use the functional and non-perturbative RG approach reviewed in~\cite{tetradis1994critical,dupuis2021the}. These two features allow us to compute results for the whole range of values of the decay parameter $\sigma$ and number $N$ of components of the field. 
A unified fRG treatment has been already carried out for both classical and quantum long-range $\mathrm{O}(N)$ models in equilibrium~\cite{defenu2020criticality}. In~\Cref{subsec:nonequilFRG} we introduce our non-equilibrium fRG setup, combining the Wetterich equation with the MSRJD description of critical dynamics with a time boundary~\cite{chiocchetta2016universal,ihssen2025nonperturbative}. The scaling of the quantities that undergo renormalization, resulting in a definition of the exponents $z$ and $\theta$, is described in~\Cref{subsec:criticalExps}. The propagators and the projection scheme used to derive the flow equations are presented in~\Cref{subsec:propagators_projections}. Finally, in~\Cref{subsec:flowEqns} we report the resulting flow equations and expressions for the dynamical critical exponents.

\subsection{Non-equilibrium fRG}\label{subsec:nonequilFRG}

Let us introduce a mass scale $k \in [0,\Lambda]$, where $\Lambda$ is a UV cutoff proportional to the inverse of the lattice spacing of~\eqref{LR Hamiltonian}. In the fRG framework we attach a $k$-dependence to the generating functional obtained from~\eqref{msr action} and related quantities via the introduction of a regulator $\mathds{R}_k$ that suppresses the propagation of low-energy modes. In particular, instead of~\eqref{msr action} one considers a description in terms of a 1PI effective average action $\Gamma_k = \Gamma_k[\boldsymbol{\phi},\tilde{\boldsymbol{\phi}}]$, which interpolates between the bare action $\Gamma_\Lambda = S$ and the (unknown) genuine effective action $\Gamma_0 = \Gamma$ as the scale $k$ is lowered. The scale-dependence of the effective action is captured by the Wetterich equation~\cite{wetterich1993exact,tetradis1994critical,dupuis2021the}, which in our non-equilibrium case ($t>t_0$) reads~\cite{chiocchetta2016universal,ihssen2025nonperturbative}
\begin{align}\label{wetterich}
	\partial_\kappa \Gamma_k[\boldsymbol{\Phi}] = \frac{1}{2} \int_{\boldsymbol{x},t>t_0} \tr \left[\mathds{G}_k[\boldsymbol{\Phi}](t,\boldsymbol{x};t,\boldsymbol{x}) \, (\partial_\kappa \mathds{R}_k) \right]\,,
\end{align}
where $\kappa = \log(k/\Lambda)$ is the RG-time, $\boldsymbol{\Phi} = (\boldsymbol{\phi},\tilde{\boldsymbol{\phi}})$ is the `superfield', the trace is over $N$-vector components and the $2\times2$ superfield structure, and
\begin{equation}\label{full propagator}
	\mathds{G}_k[\boldsymbol{\Phi}] = \left( \Gamma_k^{(2)}[\boldsymbol{\Phi}] + \mathds{R}_k \right)^{-1}\,,
\end{equation}
is the full field-dependent propagator. The regulator matrix $\mathds{R}_k$ is chosen diagonal in the $\mathrm{O}(N)$-components, while it is purely off-diagonal in $(\boldsymbol{\phi},\tilde{\boldsymbol{\phi}})$-space~\cite{canet2007non,canet2011general,dupuis2021the}. This means that we have a $2N\times2N$ block diagonal matrix, which takes the following form in momentum space 
\begin{equation}
	\mathds{R}_{k,ij}(q^\sigma) = \delta_{ij} \hat{\tau}_1 R_k(q^\sigma), \quad \hat{\tau}_1 = \begin{pmatrix}
		0 & 1
		\\
		1 & 0
	\end{pmatrix}\,,
\end{equation}
where $R_k$ -- specified later -- is assumed to be independent of time, although this may be improved as in~\cite{duclut2017frequency}. 

As usual,~\eqref{wetterich} being a functional integro-differential equation, the only possibility to proceed is to specify some ansatz for the form of the effective action. On the basis of the bare action~\eqref{msr action} a sensible choice is~\cite{chiocchetta2016universal,ihssen2025nonperturbative}
\begin{align}\label{EAA ansatz}
	& \Gamma_k = \int_{\boldsymbol{x}} \left[-\frac{Z_{0,k}^2}{2 \tau_0} \tilde{\phi}_{0,i}(\boldsymbol{x})^2 + Z_{0,k} \tilde{\phi}_{0,i}(\boldsymbol{x}) {\phi}_{0,i}(\boldsymbol{x}) \right] 
	\nonumber\\
	& + \int_{\boldsymbol{x},t > t_0} \tilde{\phi}_i (t,\boldsymbol{x}) \left[\left(Z_k \partial_t + K_k \nabla^{\sigma} \right) \phi_i(t,\boldsymbol{x}) - \Omega_k \tilde{\phi}_i (t, \boldsymbol{x}) \right]
	\nonumber\\
	& + \int_{\boldsymbol{x},t > t_0} \tilde{\phi}_i(t,\boldsymbol{x}) V_k^{(i)}(\boldsymbol{\phi}(t,\boldsymbol{x}))\,,
\end{align}
where we have introduced the (field-independent) renormalization functions $Z_{0,k},Z_k,K_k,\Omega_k$ and the fully field-dependent effective potential $V_k(\boldsymbol{\phi})$. No independent boundary renormalization is assigned to $\phi_0$. Indeed, at the initial-condition fixed point $\tau_0^{-1}=0$, its fluctuations vanish, see~\eqref{eq:correlators_initial_time}. The leading nontrivial order-parameter operator at the initial-time surface is instead $\partial_t\phi|_{t=t_0}$, which causality relates to $\tilde\phi_0$~\cite{janssen1989new,janssen1992renormalized}. Their boundary singularities are therefore absorbed by the single additional renormalization function $Z_{0,k}$ associated with $\tilde\phi_0$.

\subsection{Dimensional analysis}\label{subsec:criticalExps}

In order to uncover the fixed point solutions associated with the flows obtained from~\eqref{wetterich} (cf.~\Cref{subsec:flowEqns}), we need to first adimensionalize the quantities entering the effective action~\eqref{EAA ansatz}, which is itself dimensionless, $[\Gamma_k] = 0$. Scaling dimensions $[\cdot]$ are defined in terms of the external scale $k$. 

Since space and time scale anisotropically according to the dynamical critical exponent $z$, we consider the scaling dimension of a spatial coordinate to be $[x]=-1$, while for time $[t]=-z$. Also, we assume the typical scaling $K_k \sim k^{-\eta_K}$ of the wavefunction renormalization as $k\to0$, where $\eta_K$ is the anomalous dimension. Hence, $[K_k] = -\eta_K$, and similarly $[Z_k] = -\eta_Z$ for some $\eta_Z$ that we will determine from the flow of $Z_k$ as $\eta_Z = -\partial_\kappa \log Z_k$. From these general assumptions it follows directly that the dimensional consistency of $Z \partial_t + K \nabla^\sigma$ implies
\begin{equation}\label{definition of z}
	z = \sigma + \eta_Z -\eta_K\,,
\end{equation}
where it is known that $\eta_K = 0$ for the LR model~\cite{defenu2020criticality}. In the absence of quantum corrections, we deduce from~\eqref{definition of z} that the mean-field result is $z_{\text{mf}}=\sigma$, as anticipated. Dimensional considerations on the quadratic part of the bulk action lead to $[\phi]+[\tilde{\phi}] = d-\sigma+\eta_K+z$. Taking the dimension of $\phi$ to be the same as in equilibrium yields
\begin{equation}\label{dimension of the fields}
	[\phi] = \frac{d-\sigma+\eta_K}{2}
	\quad \text{and} \quad
	[\tilde{\phi}] = [\phi] + z\,.
\end{equation}
Moreover, if $[\Omega_k] = -\eta_\Omega$, from~\eqref{dimension of the fields} and~\eqref{definition of z} we find that $\eta_\Omega = \eta_Z$. All of this is consistent with the last line of~\eqref{EAA ansatz} provided $[V_k] = d$, the usual dimension of the effective potential. We conclude that the dimensionless and renormalized quantities to be used later are
\begin{align}
	& \overline{x} = k x\,,
	\quad
	\overline{t} = k^z t\,,
	\nonumber\\
	&\overline{\rho}(\overline{t}\,,\overline{\boldsymbol{x}}) \equiv K_k k^{\sigma-d} \rho(t,\boldsymbol{x})\,,
	\quad
	 u_k(\overline{\rho}) \equiv k^{-d} U_k(\rho)\,.
\end{align}
Let us finally consider the dimensions of the boundary action. Since $\tilde{\phi}_0$ alone is responsible for the introduction of a novel boundary exponent~\cite{janssen1989new,janssen1992renormalized,calabrese2005ageing}, we take $\phi_0 = \phi(t=0)$ to have the same dimension as $\phi$, i.e. $[\phi_0] = [\phi]$. Assuming the scaling $Z_{0,k} \sim k^{-\eta_{Z_0}}$ in the IR limit $k\to0$, from the first line of~\eqref{EAA ansatz} we find
\begin{equation}
	[\tilde{\phi}_0] = [\tilde{\phi}] + \eta_{Z_0} - \eta_Z
	\quad \text{and} \quad
	[\tau_0] = \sigma-\eta_K\,.
\end{equation}
In the absence of anomalous scaling the dimension of $\tilde{\phi}_0$ would be the same as that of the bulk response field $\tilde{\phi}$. However, in general such anomalous scaling is related to the aging exponent $\theta$ via~\cite{janssen1989new,calabrese2005ageing}
\begin{equation}\label{eq:definition of theta}
    \theta = \frac{\eta_Z-\eta_{Z_0}}{z}\,.
\end{equation}

\subsection{Propagators and projection scheme}
\label{subsec:propagators_projections} 

The derivation of the flow equations requires three main steps. We first evaluate $\Gamma_k^{(2)}+\mathds{R}_k$ on a spatially uniform background and invert it to obtain the correlation and response functions. We then take suitable field derivatives of the Wetterich equation~\eqref{wetterich} and project the resulting vertex functions onto the field, momentum, frequency, and initial-time structures retained in the ansatz~\eqref{EAA ansatz}. In the following, we outline this procedure, while the details are presented in Appendices~\ref{app:variations with boundary}, \ref{app:invert hessian}  and \ref{app:calculation of the flows}.

The $\mathrm{O}(N)$ symmetry implies that the effective potential depends on the order-parameter field only through the invariant $\rho=\boldsymbol{\phi}^2/2$, so that we write $V_k(\boldsymbol{\phi})=U_k(\rho)$. Its field derivatives are denoted by \begin{equation} \begin{aligned}\label{shorthand notation field derivatives of potential} \partial_{\phi_i}V_k &\equiv V_k^{(i)} =\phi_iU_k'(\rho),\\ \partial_{\phi_i}\partial_{\phi_j}V_k &\equiv V_k^{(ij)} =\delta_{ij}U_k'(\rho) +\phi_i\phi_jU_k''(\rho), \end{aligned} \end{equation} and analogously for higher derivatives.

To evaluate the Hessian and the vertex functions, we expand around a spatially uniform background, \begin{subequations}\label{uniform background} \begin{align} \boldsymbol{\phi}_*(t) &= \boldsymbol{\phi}_u+\delta\boldsymbol{\phi}(t), &\boldsymbol{\phi}_u &= \bigl(\sqrt{2\rho},0,\ldots,0\bigr),\\ \tilde{\boldsymbol{\phi}}_*(t) &= \tilde{\boldsymbol{\phi}}_u+\delta\tilde{\boldsymbol{\phi}}(t), &\tilde{\boldsymbol{\phi}}_u &= \boldsymbol{0}. \end{align} \end{subequations}
The orientation of $\boldsymbol{\phi}_u$ is chosen without loss of generality by virtue of the $\mathrm{O}(N)$ symmetry, while the auxiliary response field vanishes on the physical background in the absence of external sources~\cite{canet2011general}. For the bulk projections, the time-dependent corrections can be neglected and the expansion is performed around the stationary configuration $\boldsymbol{\Phi}_u=(\boldsymbol{\phi}_u,\boldsymbol{0})$. They must instead be retained when extracting contributions localized at the initial-time surface. After taking the required functional derivatives, we evaluate the resulting expressions at the running minimum $\rho=\rho_{\min,k}$ of the effective potential, defined by $U_k'(\rho_{\min,k})=0$.

On the stationary background $\boldsymbol{\Phi}_u$, the regulated inverse propagator, see~\eqref{full propagator}, is diagonal in the $\mathrm{O}(N)$ indices. The direction selected by $\boldsymbol{\phi}_u$ defines a longitudinal mode, while the remaining $N-1$ components correspond to transverse modes. Their field-dependent renormalized dispersions are \begin{subequations}\label{renormalized dispersions main} \begin{align} 
    \omega_{L,k}(q^\sigma) &= K_k q^\sigma+R_k(q^\sigma) +U_k'(\rho)+2\rho U_k''(\rho),\\ \omega_{T,k}(q^\sigma) &= K_k q^\sigma+R_k(q^\sigma) +U_k'(\rho). 
\end{align} 
\end{subequations}
Accordingly, the inversion of $\mathds{G}_k^{-1}$ reduces to the separate inversion of one longitudinal and $N-1$ identical transverse $2\times2$ blocks in correlation-response field space.

In the bulk (equivalently, in the limit $t_0\to-\infty$) time-translation invariance is restored and the regulated Hessian can be inverted in frequency space. Transforming back to the time domain gives, for each sector $i\in\{L,T\}$, 
\begin{equation}\label{equilibrium propagator matrix}
	\mathds{G}_{\text{eq},i}(t,t') = \begin{pmatrix}
		G^C_{\text{eq},i}(t,t') & G^R_{\text{eq},i}(t,t') \\
		G^R_{\text{eq},i}(t',t) & 0
	\end{pmatrix}\,,
\end{equation}
where
\begin{subequations}\label{equilibrium correlators}
	\begin{align}
		&G^C_{\text{eq},i}(t,t') = \frac{\Omega_k}{Z_k \omega_{i,k}} \left[e^{-\omega_{i,k}|t-t'|/Z_k} \right]\,,
		\label{equilibrium correlation function}
		\\
		&G^R_{\text{eq},i}(t,t') = \frac{\vartheta(t-t')}{Z_k} e^{-\omega_{i,k}(t-t')/Z_k}\,.
		\label{equilibrium response function}
	\end{align}
\end{subequations}
These time-translation-invariant propagators are used to determine the bulk flows of $U_k$, $K_k$, $Z_k$, and $\Omega_k$. Their explicit derivation is presented in~\Cref{app:invert hessian}. Here and in what follows, the subscript `eq' (for `equilibrium') indicates that the propagator is time-translation invariant, depending on $t$ and $t'$ only through $t-t'$.

To determine the flow of the boundary renormalization function $Z_{0,k}$, the initial time $t_0$ must instead be kept finite, which breaks time-translation invariance. The correlation function $G^C_{0,i}(t,t') = G^C_{\mathrm{eq},i}(t,t') + G^C_{\mathrm{neq},i}(t,t')$ acquires an additional transient contribution
\begin{equation}\label{nonequilibrium correlation main} 
    G^C_{\mathrm{neq},i}(t,t') \propto e^{-\omega_{i,k}(t+t'-2t_0)/Z_k}.
\end{equation}
In contrast, the response function remains unchanged, $G^R_{0,i}(t,t')=G^R_{\mathrm{eq},i}(t,t')$. Although the additional term in~\eqref{nonequilibrium correlation main} decays exponentially away from the initial-time surface, it must be retained to extract the boundary-localized flow of $Z_{0,k}$. The full expression for its prefactor and the inversion of the Hessian at finite $t_0$ are given in~\Cref{app:invert hessian}.

Having obtained the propagators, we take suitable field derivatives of the Wetterich equation. Denoting either $\boldsymbol{\phi}$ or $\tilde{\boldsymbol{\phi}}$ collectively by $\boldsymbol{\Phi}$, the variation of the full propagator follows from 
\begin{equation} 
    \delta_{\boldsymbol{\Phi}}\mathds{G}_k = -\mathds{G}_k \left(\delta_{\boldsymbol{\Phi}}\Gamma_k^{(2)}\right) \mathds{G}_k\,. 
\end{equation} 
A second field derivative consequently generates the standard sunset contributions, containing three propagators and two first variations of the Hessian, and a tadpole contribution, containing two propagators and one second variation of the Hessian. Products in these expressions imply integrations over intermediate spacetime arguments and summations over internal indices. The explicit functional variations, including the contributions localized at $t=t_0$, are given in~\Cref{app:variations with boundary}.

The running functions are identified by matching the field derivatives of the Wetterich equation with the corresponding structures in the ansatz~\eqref{EAA ansatz}. The flow of the potential is obtained from the one-point response-field vertex on the uniform background, \begin{equation} \partial_\kappa U_k'(\rho) = \left. \frac{1}{\sqrt{2\rho}}\, \frac{\delta}{\delta\tilde{\phi}_1} \partial_\kappa\Gamma_k \right|_{\boldsymbol{\Phi}=\boldsymbol{\Phi}_u}, \end{equation} where vanishing external momentum and frequency are understood. For the remaining bulk functions, we take two derivatives in the transverse sector $i=g\neq1$ (see the discussion below). Defining the corresponding mixed two-point flow by \begin{equation} \partial_\kappa\Gamma_{k,\tilde{\phi}_g\phi_g}^{(2)} (\omega,\boldsymbol{q}) \equiv \left. \frac{\delta^2\partial_\kappa\Gamma_k} {\delta\tilde{\phi}_g(-\omega,-\boldsymbol{q})\, \delta\phi_g(\omega,\boldsymbol{q})} \right|_{\boldsymbol{\Phi}=\boldsymbol{\Phi}_u}, \end{equation} the flows of $K_k$ and $Z_k$ are determined by \begin{align} \partial_\kappa K_k &= \left. \partial_{q^\sigma} \partial_\kappa\Gamma_{k,\tilde{\phi}_g\phi_g}^{(2)} (0,\boldsymbol{q}) \right|_{\boldsymbol{q}=0}, \\ \partial_\kappa Z_k &= \left. -i \partial_{\omega} \partial_\kappa\Gamma_{k,\tilde{\phi}_g\phi_g}^{(2)} (\omega,\boldsymbol{0}) \right|_{\omega=0}. \end{align} Since the loop corrections do not generate a nonanalytic contribution proportional to $q^\sigma$, the first of these projections vanishes, so $\partial_\kappa K_k = 0$. The noise amplitude is obtained from the response-response two-point function, \begin{equation} \partial_\kappa\Omega_k = -\frac{1}{2} \left. \frac{\delta^2\partial_\kappa\Gamma_k} {\delta\tilde{\phi}_g(-\omega,-\boldsymbol{q})\, \delta\tilde{\phi}_g(\omega,\boldsymbol{q})} \right|_{\substack{ \boldsymbol{\Phi}=\boldsymbol{\Phi}_u\\ \omega=0,\ \boldsymbol{q}=0}}. \end{equation} Finally, $Z_{0,k}$ is extracted at finite $t_0$ by matching the coefficient of the boundary-localized variation $\delta\phi_{0,g}\delta\tilde{\phi}_{0,g}$: \begin{equation} \partial_\kappa Z_{0,k}\, \delta\phi_{0,g}\delta\tilde{\phi}_{0,g} = \left. \frac{\delta^2\partial_\kappa\Gamma_k} {\delta\phi_g\,\delta\tilde{\phi}_g} \right|_{\boldsymbol{\Phi}=\boldsymbol{\Phi}_u} \bigg|_{\delta\phi_{0,g}\delta\tilde{\phi}_{0,g}}. \end{equation}

For the bulk projections, we retain the time-translation-invariant part of the propagator and discard the transient contributions, which are exponentially suppressed with the distance from the initial-time surface. The boundary projection is treated separately: the initial time is kept finite and the contribution proportional to $\delta\phi_{0,g}\delta\tilde{\phi}_{0,g}$ is isolated using the short-time expansion described in~\Cref{app:trick_short_time}. Applying these projections yields the flow equations and anomalous dimensions summarized in the following subsection, while their explicit evaluation is presented in~\Cref{app:calculation of the flows}.

\subsection{Flow equations and critical exponents}\label{subsec:flowEqns}

Applying the projections described above, we obtain the following flow equation for the derivative of the dimensionless effective potential:
\begin{align}\label{flow eqn derivative effective potential}
    &\partial_\kappa u_k'(\overline{\rho}) = (-\sigma+\eta_K) u_k'(\overline{\rho})
    + (d-\sigma+\eta_K) \overline{\rho} u_k''(\overline{\rho})
	\\
	&- \frac{4 v_d}{\sigma} \left[ \mu_L'(\overline{\rho}) L_1^{(d,\sigma)}(\mu_L) + (N-1) {\mu}_T'(\overline{\rho}) L_1^{(d,\sigma)}({\mu}_T) \right]\,,
	\nonumber
\end{align}
where $v_d^{-1} = 2^{d+1} \pi^{d/2} \Gamma(d/2)$, the functions $L_n^{(d,\sigma)}$ are described in~\Cref{subsec:threshold functions}, and the longitudinal and transverse masses are
\begin{align}
	&\mu_L(\overline\rho) = u_k'(\overline\rho) + 2 \overline\rho u_k''(\overline\rho)
	\quad \text{and} \quad
	\mu_T(\overline\rho) = u_k'(\overline\rho)\,.
\end{align}
Equation~\eqref{flow eqn derivative effective potential} is consistent with static fRG~\cite{defenu2015fixed,defenu2020criticality,defenu2023long}. The dimensionless form is particularly useful because, instead of studying the full RG flow of the potential, we can focus on its fixed-point value $u'_*(\rho)$, obtained by setting $\partial_\kappa u'_k = 0$ and solving the resulting ordinary differential equation. Physically, this captures precisely the scaling properties relevant to the critical behavior.

For each function $A_k \in \{ Z_{0,k}, Z_k,K_k,\Omega_k\}$ we define the corresponding anomalous dimensions $\eta_A \coloneqq -\partial_\kappa \log A_k$. As mentioned above, $K_k$ does not flow, and therefore $\eta_K=0$. Moreover, we observe $\eta_\Omega = \eta_Z$, which corroborates \textit{a posteriori} our dimensional analysis. The explicit expressions for $\eta_{Z_0}$ and $\eta_Z=\eta_\Omega$ are given in Eqs.~\eqref{etaZ0 complete goldstone} and~\eqref{goldstone etaZ}, respectively. For a generic regulator $R_k(q^\sigma)$, they take the form of momentum integrals involving derivatives of the effective potential. Together with Eqs.~\eqref{eq:definition of theta} and~\eqref{definition of z}, they determine the dynamical exponents $\theta$ and $z$. From a technical point of view, we obtain~\eqref{etaZ0 complete goldstone} and~\eqref{goldstone etaZ} upon projecting to one of the transverse components $i \neq 1$, as discussed in~\cite{ihssen2025nonperturbative}.

The present calculation corresponds to the leading iteration of the expansion introduced in Ref.~\cite{ihssen2025nonperturbative}. At this order, the bulk couplings and vertices are obtained from the fully interacting, time-translation invariant fixed-point solution, while the propagator retains the finite-$t_0$ contribution required for the boundary projection. This already generates an independent flow of $Z_{0,k}$ and hence the nontrivial aging exponent~\eqref{eq:definition of theta}. The bulk projections are instead performed onto the time-translation-invariant sector, where detailed balance implies $\eta_\Omega=\eta_Z$ and determines the asymptotic exponent $z$. Further iterations would retain the explicitly time-dependent corrections generated by the flow and feed them back into the propagator. Promoting $Z_k$ and $\Omega_k$ to time-dependent functions would then allow one to capture transient deviations from the equilibrium fluctuation-dissipation relation. Such corrections vanish as the system relaxes toward equilibrium, but may provide quantitative corrections to the finite-time dynamics.

The numerical results for the dynamical exponents in~\eqref{definition of z} and~\eqref{eq:definition of theta} are obtained by substituting the fixed-point $u_*(\overline{\rho})$ of the effective potential and its derivatives evaluated at the minimum $\overline{\rho}_{\text{min}}$. In fact, it is straightforward to make~\eqref{etaZ0 complete goldstone} and~\eqref{goldstone etaZ} dimensionless, so that at the critical point they exhibit no $k$-dependence, with only the dimensionless potential entering these expressions.

\section{Results}\label{sec:results}

In this section we present the results obtained for the critical exponents $\theta$ and $z$ describing the dynamical scaling properties of a long-range model subject to a sudden critical quench at time $t=t_0$. We have obtained the fixed-point potential $u_*(\overline{\rho})$ coming from our fRG analysis by making an explicit regulator choice:
\begin{equation}\label{Litim regulator}
	R_k(q^{\sigma}) = K_k (k^\sigma-q^\sigma) \, \theta (k^\sigma-q^\sigma)\,,
\end{equation}
which is a generalization of the Litim regulator~\cite{litim2000optimization, litim2001optimized} suitable to the long-range case~\cite{defenu2015fixed}. This form of the cutoff function allows for an analytic integration of the threshold functions, which are explicitly given in~\Cref{subsec:threshold functions}. The numerical solution of the differential equation for the effective potential has been obtained using a combination of a shooting approach and a pseudo-spectral collocation method; the details are found in~\cite{ihssen2025nonperturbative}. We also remark that, as opposed to the latter study about short-range models, in the long-range case $\eta_K$ vanishes, thus simplifying the computational procedure.

\subsection{One-dimensional Ising model}\label{subsec:1DLRI}

\begin{figure}
    \centering
    \includegraphics[width=0.9\linewidth]{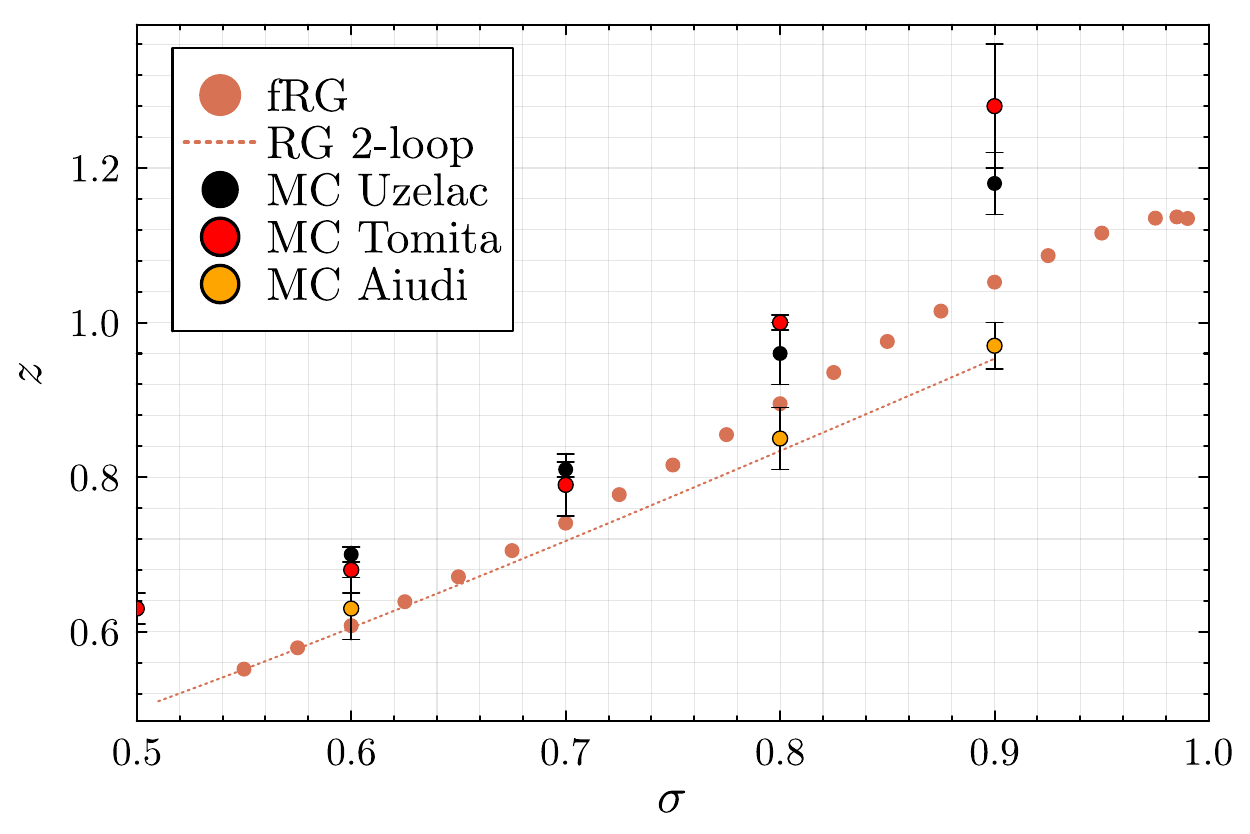}
    \includegraphics[width=0.9\linewidth]{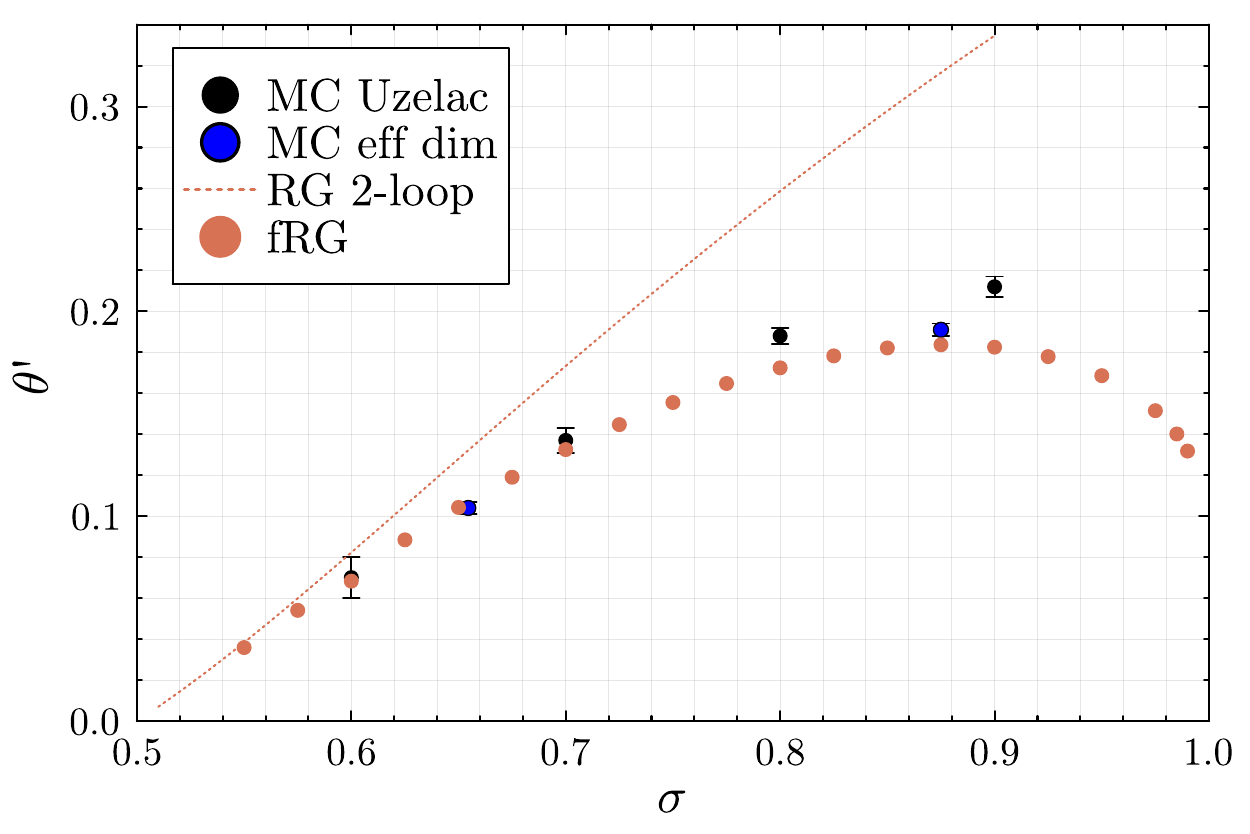}
    \caption{Dynamical exponents $z$ (upper panel) and $\theta'$ (lower panel) for the long-range Ising model in $d=1$. The dark orange dots are obtained using the fRG scheme described in~\Cref{sec:frg}, with the regulator~\eqref{Litim regulator}. The dotted lines show the weak-coupling expansions of~\cite{chen2000short}, which is only valid in the vicinity of $\sigma = 0.5$. The dots with error bars are MC estimates from Ref.~\cite{uzelac2008short} (black), Ref.~\cite{tomita2008monte} (red), the dynamical Lévy lattice simulations of Ref.~\cite{aiudi2023critical} (yellow),  and, as explained in the main text,  Refs.~\cite{grassberger1995damage,nightingale2000monte,hasenbusch2020dynamic} via the effective dimension equivalence (blue).\hspace*{\fill}}
    \label{fig:one-dimension}
\end{figure}

We begin with a discussion of results in $d=1$, due to the existence of Monte Carlo (MC) simulations of the long-range Ising model in one spatial dimension~\cite{uzelac2008short,tomita2008monte}. In~\Cref{fig:one-dimension} we show the dynamical exponent $z$ and the initial slip exponent $\theta'$, defined in~\eqref{magnetization initial slip}, for several values of the range parameter $\sigma \in (\sigma_{\rm mf}=1/2,\sigma_*=1)$. For $z$, we find that our result lies in between the MC points of~\cite{uzelac2008short,tomita2008monte} and the two-loop formula obtained from the perturbative RG in~\cite{chen2000short}. The latter is in good agreement for small $\sigma$, which confirms the consistency of our non-perturbative approach in the weak-coupling limit. In addition,~\Cref{fig:one-dimension} displays MC values obtained through the effective dimension approach of~\Cref{sec:effdim}, as explained below. We note that, as discussed in more detail in~\cite{ihssen2025nonperturbative}, the accuracy of the present fRG computation can be improved with different regulator choices and extensions of the expansion scheme.

A further, independent benchmark is provided by Ref.~\cite{aiudi2023critical}, where the dynamical exponent is extracted from the critical relaxation time of the long-range Ising chain simulated on a dynamical Lévy lattice, a local algorithm whose validity is established by reproducing the known static exponents. Their estimates are shown in the upper panel of~\Cref{fig:one-dimension}. They lie systematically below the values of Refs.~\cite{uzelac2008short,tomita2008monte} and close to the two-loop prediction of~\cite{chen2000short}, so that for large $\sigma$ our fRG curve is bracketed from below and from above by independent Monte Carlo results.

In the lower panel of~\Cref{fig:one-dimension}, we plot our results for the initial slip exponent, which are in good agreement with the MC points of \cite{uzelac2008short}, as well as those coming from the effective dimension approach. Especially for this non-equilibrium critical exponent, a remarkable improvement over the perturbative RG (red dotted curve) is apparent. 

We remark that, for both exponents $z$ and $\theta'$, the correct behavior as $\sigma \to \sigma_* = 1$ is not known \textit{a priori}. To the best of our knowledge, no previous studies have addressed this regime, and our results therefore constitute the first estimates of the dynamical exponents near the short-range crossover. It is worth noting, however, that even in equilibrium, a similar truncation of the effective action within the fRG framework correctly reproduces the limiting value of the exponent $\nu$ as $\sigma \to 1$, but not its approach to this limit, see~\cite{pagni2025one}. The delicate nature of this region in $\sigma$-space is related to the fact that the one-dimensional short-range Ising model has no finite-temperature transition. In contrast, in~\Cref{subsec:twodimensional_results} we show that for $d=2$ our approach correctly captures the crossover to short-range universality as $\sigma \to \sigma_*$.

Finally, let us detail how to apply the effective dimension approach of~\Cref{sec:effdim} to MC data obtained for short-range Ising models in two and three dimensions. In~\cite{grassberger1995damage} we find $(\theta')_{\text{SR}}^{\text{MC}} = 0.191(3)$ for $D=2$ and $(\theta')_{\text{SR}}^{\text{MC}} = 0.104(3)$ for $D=3$. For the dynamical critical exponents there exist several MC studies: in $D=2$ it is found in~\cite{nightingale2000monte} that $z_{\text{SR}}^{\text{MC}} = 2.1667(5)$, while $z_{\text{SR}}^{\text{MC}}= 2.0245(15)$ in $D=3$ is reported in~\cite{hasenbusch2020dynamic}. These results are obtained at a fixed value of the (short-range) dimension $D$. 
Using the correspondences ~\eqref{effective dimension z} and~\eqref{effective dimension theta}, we can compute the LR equivalent of these quantities 
\begin{subequations}\label{mapping MC}
\begin{align}
	z_{\text{LR}}^{\text{MC}}(\sigma) &\equiv \frac{d}{D} z_{\text{SR}}^{\text{MC}}(D)\,,
    \label{mapping MC a}
	\\
	{\theta'}_{\text{LR}}^{\text{MC}}(\sigma) &= {\theta'}_{\text{SR}}^{\text{MC}}(D) \,,
    \label{mapping MC b}
\end{align}
\end{subequations}
where $d=1$, and the two values of $\sigma$ are obtained from $D=2$ and $D=3$ through~\eqref{effective dimension}:
\begin{equation}\label{effective sigma}
	\sigma = (2-\eta_{\text{SR}}(D)) \frac{d}{D}\,.
\end{equation}
Here $\eta_{\rm SR}(2) = 1/4$, while $\eta_{\rm SR}(3) = \eta_{\rm CB} \approx 0.0363$ is taken from the recent conformal bootstrap (CB) work \cite{chang2025bootstrapping}. Therefore, the blue dots of~\Cref{fig:one-dimension} do not rely on the results of the fRG: This provides an \textit{independent} verification that the effective dimension approach yields estimates of the dynamical exponents that -- by visual inspection of the curves in~\Cref{fig:one-dimension} -- are in qualitative agreement with the MC results for the long-range Ising chain. However, as discussed in \Cref{sec:effdim}, we do not expect that the dimensional correspondence is exact.

\subsection{Long-range critical heat engine}

Before a more complete discussion of the dynamical critical $O(N)$ behavior in higher dimensions, here we would like to propose long-range interactions as a possible way to obtain a thermodynamic advantage in the operations of a so-called \textit{critical heat engine}.

Traditionally, it has been thought that reaching the Carnot efficiency $\eta_{\rm C}$ of a heat engine implies working in the quasi-static limit, i.e.~with zero power output $\mathcal{P}$. More generally, there is a trade-off between the power $\mathcal{P}$ and the efficiency $\eta$ of a thermodynamic heat engine~\cite{pietzonka2018universal}. A way to optimize the ratio $\dot{\Pi} \equiv \mathcal{P}/(\eta_{\rm C}-\eta)$, which we refer to as `performance rate', was proposed in Ref.~\cite{campisi2016power}: One can design a thermodynamic Otto cycle where -- instead of a non-interacting substance -- the working medium is a system close to its critical point, so that the scaling of the performance rate with the size $N$ of the medium is given by
\begin{equation}
    \dot{\Pi} \sim N^{1+\frac{\pi_{\rm th}}{d \nu}}\,,
\end{equation}
where $\pi_{\rm th} = \alpha-z\nu$ and $\alpha$ is the specific-heat exponent. Intuitively, the static part comes from the fact that the work output of a single cycle of the engine can be increased by enhancing the specific heat $c \sim |T-T_c|^{-\alpha}$ of the working substance, while operating the engine in finite time requires that the duration of the cycle is at least equal to the relaxation time $\tau_{\rm relax} \sim \xi^z \sim |T-T_c|^{-z \nu}$.

In order to increase the performance rate $\dot{\Pi}$, one aims at the maximization of the exponent $\pi_{\rm th}$. Using long-range systems provides a way of doing so, at least with respect to short-range interacting ones. In fact, a key observation of the present work is that \textit{long-range interactions generically facilitate relaxation close to criticality} by improving the scaling of $\tau_{\rm relax}$. This may be thought of as a speed-up of communication across the system, as a consequence of the enhancement in the cooperation between spins due to their long-range interactions. We quantify this effect via the dynamical exponent $z$, which already at the mean-field level, $z_{\rm mf}=\sigma$, can be significantly lower than those for the corresponding short-range systems, where we have the rigorous bound $z \geq 2$~\cite{masaoka2025rigorous}. The correlations captured by the renormalization group are consistently bringing a positive correction to the value of $z$, as seen in \Cref{fig:one-dimension} and \ref{fig:dynamical_exponent_all}. The generality of these observations is corroborated by looking at other systems with long-range interactions, e.g.~the long-range Ising model with random impurities, exhibiting an exponent $z = \sigma + O(\sqrt{\epsilon})$ \cite{chen2002short}, with $\epsilon = 2\sigma-d$, smaller than that of the nearest-neighbor random Ising model (note, however, that introducing quenched disorder tends to have the effect of slowing down relaxation with respect to the pure case).

It is noteworthy that \cite{campisi2016power} takes into account the possibility of critical speeding-up, characterized by $z < 0$. Such behavior is mostly associated either with certain Monte Carlo dynamics \cite{deng2007critical,deng2007dynamic,elcci2013efficient}, which, however, do not appear to correspond to physical stochastic evolutions, as classified by Hohenberg and Halperin \cite{hohenberg1977theory}, or with systems exhibiting unconventional relaxation mechanisms, see e.g. \cite{zappoli1990anomalous,boukari1990critical,grams2014critical,tavora2014quench}. Nonetheless, it remains plausible that the presence of long-range interactions in such systems could further accelerate thermalization, depending on the specific microscopic pathways to relaxation. Notably, in Ref. \cite{grams2014critical}, where $z<0$ is experimentally observed at the monopole liquid-gas transition in a spin-ice compound, long-range Coulomb interactions between emergent monopoles are already intrinsic to the system.

\begin{figure}
    \centering
    \includegraphics[width=\linewidth]{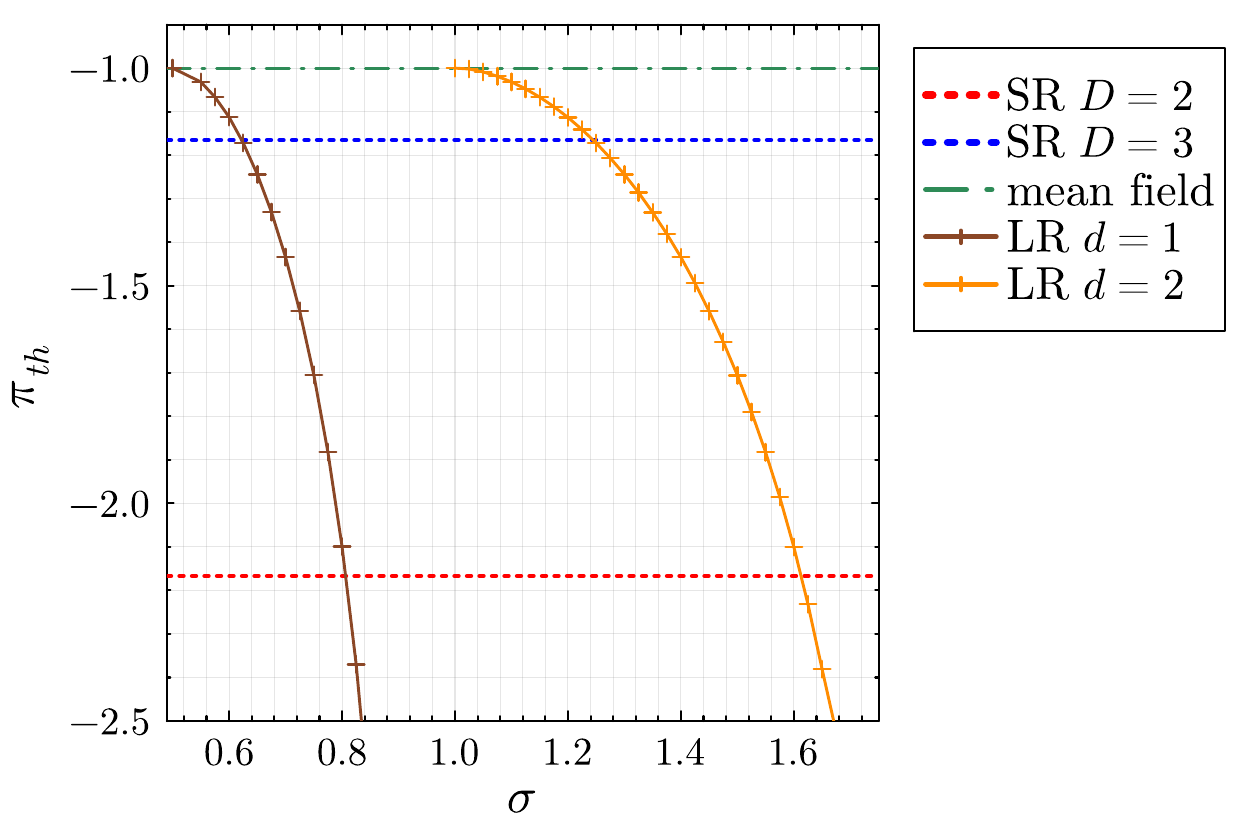}
    \caption{Performance-rate exponent $\pi_{\rm th} = \alpha-z\nu$ for long-range (LR) and short-range (SR) Ising models. The horizontal red and blue dashed lines correspond to the SR Ising model in dimension $D=2$ and $D=3$, respectively. The brown and yellow lines represent the values of the exponent $\pi_{\rm th}$ as $\sigma$ is varied for the LR Ising model in dimension $d=1$ and $d=2$, respectively. The latter curves are obtained by interpolating fRG data points, shown as crosses. The mean-field value $\pi_{\rm th} = -1$, reached by all long-range models at $\sigma_{\rm mf}=d/2$, is visualized as a horizontal green dash-dotted line.\hspace*{\fill}}
\label{fig:performance_rate_exponent}
\end{figure}

On the other hand, the specific-heat exponent $\alpha$ is also generally reduced by the presence of long-range interactions, thus implying that $\pi_{\rm th}$ is not enhanced in both its static and dynamical parts. However, we observe an overall advantage. Using the hyperscaling relation $\alpha = 2-d \nu$, which holds (within error bars) in the MC simulation \cite{tomita2008monte} of the long-range Ising chain and for self-avoiding Lévy flights \cite{sarkar2025long} in the region $\sigma_{\rm mf}<\sigma<\sigma_{*}$, enables the usage of the values $\nu(\sigma)$ obtained by some of us for $d=1$ \cite{pagni2025one} and $d=2$ \cite{defenu2015fixed}. Hence, we show in \Cref{fig:performance_rate_exponent} that there are intervals of values of $\sigma$ where the performance-rate exponent $\pi_{\rm th} = 2-(d+z)\nu$ is larger than those of the two- and three-dimensional Ising model with short-range interactions. The latter are calculated according to the values reported in Refs. \cite{nightingale2000monte,hasenbusch2020dynamic,chang2025bootstrapping}.

In particular, we notice that $\pi_{\rm th}(\sigma)$ reaches values larger than $\pi_{\rm th}^{\rm SR}(D=3) \approx -1.17$ as soon as $\sigma$ is sufficiently close to $\sigma_{\rm mf} = d/2$, where, using the mean-field behavior $z=\nu^{-1}=\sigma$, the dimension-independent result $\pi_{\rm th} = -1$ is obtained. We conclude that long-range interactions with $\sigma$ close to the mean-field limit yield an enhancement in the scaling of the performance rate $\dot{\Pi}$. This finding aligns with Refs.~\cite{solfanelli2023quantum,solfanelli2025universal}, which highlight the thermodynamic advantages of long-range interactions in \textit{quantum} many-body systems. Of course, in the present analysis the possible quantum nature of the system is irrelevant since the heat engine operates at finite temperatures where universal behavior corresponds with the one of the classical theory.

Some remarks are in order: Although the analysis in~\cite{campisi2016power} mainly addresses the mean work output of the heat engine, the presence of a critical working medium also introduces substantial fluctuations in the work output. These fluctuations pose significant challenges for the practical implementation of macroscopic critical heat engines. However, they can be mitigated in the mesoscopic regime, where finite-size effects regulate critical behavior. In this regime, it becomes possible to design heat engines that simultaneously achieve high power output and large efficiency~\cite{holubec2017work}.

More generally, it is desirable to identify working media whose thermodynamic properties exhibit scaling behavior while maintaining a reduced level of fluctuations. In this regard, the phenomenology of short-time universal dynamics—and in particular the role of the exponent $\theta$—may prove decisive for future implementations of finite-time critical heat engines. Furthermore, the presence of long-range interactions provides an additional and versatile means to control the system’s dynamical evolution. For instance, as discussed in \Cref{subsec:twodimensional_results}, tuning the crossover time $t_{\rm cross}$ offers a powerful mechanism to foster equilibration in the system.

\subsection{Two-dimensional $O(N)$ models}\label{subsec:twodimensional_results}

So far, we have considered the long-range Ising model, corresponding to the case $N=1$ of the $O(N)$-symmetric field theory \eqref{phi4 action}. We now extend the analysis to generic $O(N)$ models with $N \geq 1$ in $d=2$, enabling a discussion of the crossover to short-range interactions and a systematic comparison with short-range models through the effective dimension approach of \Cref{sec:effdim}.
\begin{figure}
	\centering
	\includegraphics[width=1.0\linewidth]{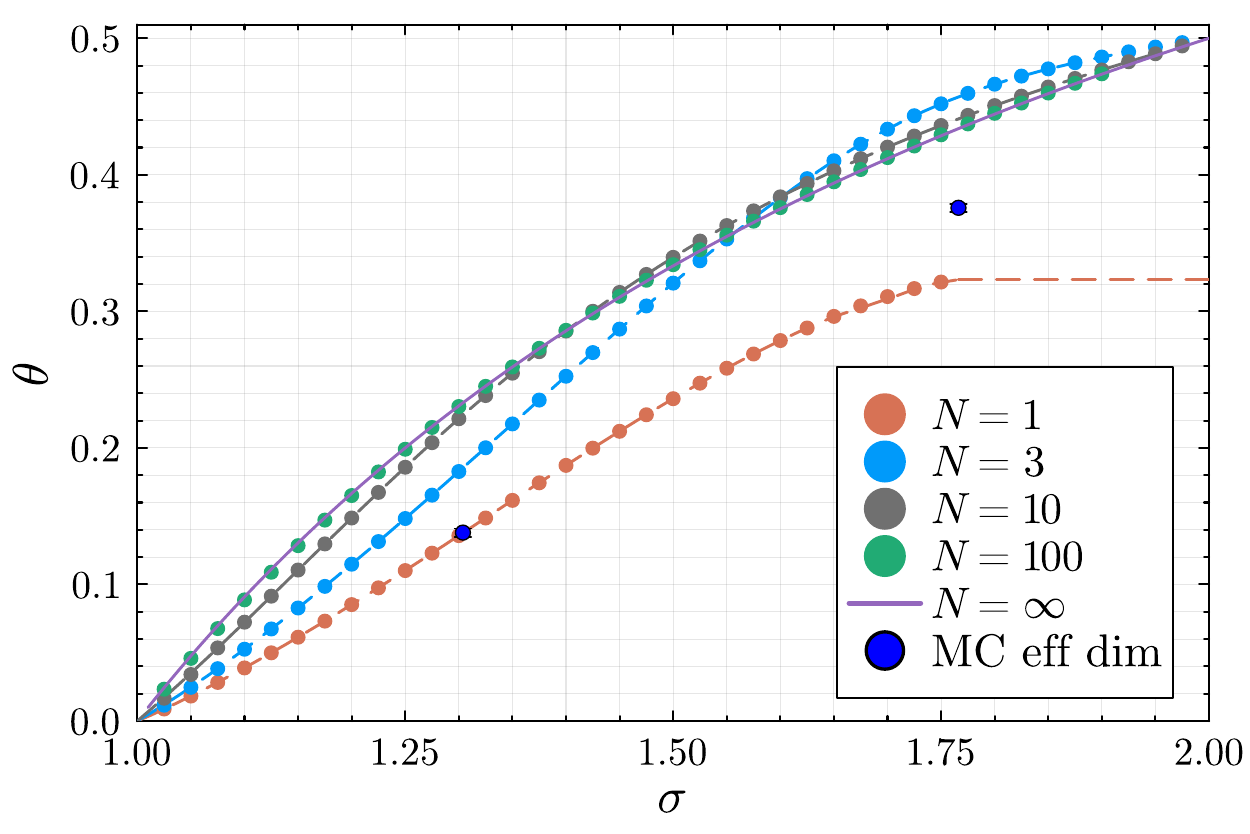}
	\caption{Aging exponent $\theta = \theta(\sigma)$ for the long-range $\mathrm{O}(N)$ models in $d=2$ with $N=1,3,10,100$. The colored dots represent data points obtained via the fRG scheme. The large-$N$ limit~\eqref{aging exponent largeN} is denoted by the solid violet line. The dashed lines are $\theta_{\text{SR}}(D(\sigma))$ for $N=1,3,10$ obtained from the short-range model through the effective dimension approach of~\Cref{sec:effdim}. The horizontal dashed line for $N=1$ starting at $\sigma \approx 1.75$ shows the role of the short-range term in the Ising case for large values of $\sigma$. The blue dots are the effective dimension Monte Carlo estimates for $N=1$, obtained from the short-range model via~\eqref{mapping MC a}, \eqref{mapping MC c} and~\eqref{effective sigma}.\hspace*{\fill}}
	\label{fig:slip_exponent_all}
\end{figure}

In~\Cref{fig:slip_exponent_all} we show the aging exponent $\theta = \theta(\sigma)$ for several values of $N$. The range of the decay parameter is $\sigma_{\text{mf}}=1 < \sigma < \sigma_*(N)$, where $\sigma_*$ is the Sak boundary introduced in~\Cref{sec:model}. For $N>1$, one has $\sigma_*(N) = 2$. In fact, in this case the theory possesses continuous symmetry and the corresponding short-range model in two dimensions cannot have spontaneous symmetry breaking due to the Mermin-Wagner theorem~\cite{mermin1966absence}, implying $\eta_\text{SR} = 0$. For $N=1$, instead, $\sigma_*(1) = 2 -\eta_{\text{SR}}$, where $\eta_{\text{SR}}=1/4$ is the anomalous dimension of the two-dimensional Ising model. As is well-known~\cite{defenu2015fixed,defenu2020criticality,solfanelli2024universality}, the competition between long- and short-range effects becomes relevant only very close to $\sigma_*(1)$. For this reason, we restrict our analysis to $\sigma \leq \sigma_*(1)$; for all $\sigma>\sigma_*(1)$ one recovers the short-range value $\theta(\sigma)\equiv\theta_{\text{SR}}$. In the opposite regime $\sigma \to \sigma_{\text{mf}} = 1$ all our curves converge to the mean-field result $\theta_{\text{mf}} = 0$ anticipated in~\Cref{sec:model}. 

We observe a rapid, and uniform in $\sigma$, convergence of the aging exponent $\theta$ towards its large-$N$ limit~\eqref{aging exponent largeN} as $N$ increases. Defining $\sigma_{\text{int}}(N)$ as the intersection point between the finite-$N$ and large-$N$ curves, we find from \Cref{fig:slip_exponent_all} that the approach is from below for $\sigma < \sigma_{\text{int}}$ and from above otherwise, with $\sigma_{\text{int}}$ decreasing as $N$ grows. The case $N=100$ is  indistinguishable from the $N\to\infty$ limit~\footnote{We note that integrating the fixed-point equation for large $N$ and $\sigma \to 2$ becomes numerically challenging.}. These benchmarks confirm the robustness of our approach and, in particular, of the truncation~\eqref{EAA ansatz}.
   
\begin{figure}
	\centering
	\includegraphics[width=1.0\linewidth]{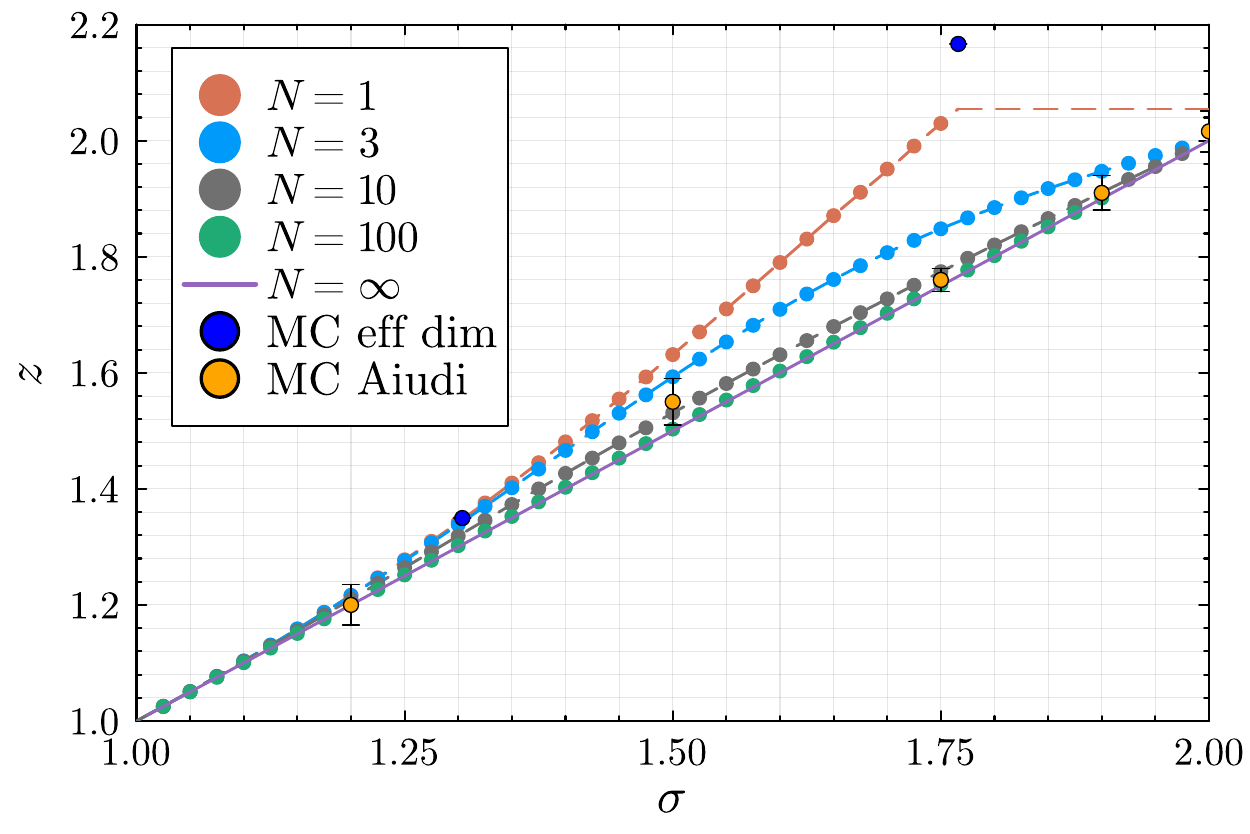}
	\caption{Dynamical exponent $z = z(\sigma)$ for the long-range $\mathrm{O}(N)$ models in $d=2$ with $N=1,3,10,100$. The colored dots represent data points obtained via the fRG scheme. The large-$N$ limit $z=\sigma$ is denoted by the solid violet line. The dashed lines are $z_{\text{SR}}(D) \, d / D$ -- cf.~\eqref{effective dimension z} and~\eqref{effective dimension} -- for $N=1,3,10$ obtained from the short-range model through the effective dimension approach. The horizontal dashed line for $N=1$ has the same meaning as in~\Cref{fig:slip_exponent_all}. The blue dots are effective dimension Monte Carlo estimates for $N=1$. The yellow dots are the direct long-range MC estimates for $N=1$ obtained on a dynamical Lévy lattice in Ref.~\cite{aiudi2023critical}.\hspace*{\fill}}
	\label{fig:dynamical_exponent_all}
\end{figure}

Likewise, we show in~\Cref{fig:dynamical_exponent_all} the dynamical exponent $z$ for $d=2$ and various $N$.
Again, the mean-field limit $z_{\text{mf}} = \sigma$ is reached -- independent of $N$ -- as $\sigma \to \sigma_{\text{mf}}=1$. Similarly, in the short-range limit $\sigma \to \sigma_*$ we retrieve $z = z_{\text{SR}}(N=1)>2$ for $N=1$ and $z=2$ for $N>1$, in agreement with the Mermin-Wagner theorem. The large-$N$ limit $z = \sigma$ is also approached uniformly from above as $N$ grows larger.

Moreover, our results, obtained by working directly with the long-range model, match almost perfectly with the short-range ones once the effective dimension approach described in~\Cref{sec:effdim} is employed. Indeed, we have used the data obtained in Ref.~\cite{ihssen2025nonperturbative} on the critical exponents in fractional dimensions in the range $D\in(2,4)$ to determine the effective dimension~\eqref{effective dimension} as a continuous function of $\sigma$. The dashed curves in~\Cref{fig:slip_exponent_all} and~\ref{fig:dynamical_exponent_all}, which overlap almost perfectly with the data points, are not merely guides to the eye: they were obtained independently, that is, via the effective-dimension relations~\eqref{effective dimension},~\eqref{effective dimension z},~\eqref{effective dimension theta} using purely short-range fRG data. In fact, the dimensional equivalence can be observed already at the level of the fRG equations, which can be mapped to their short-range counterparts~\cite{ihssen2025nonperturbative}. This gives further support to our arguments in~\Cref{sec:effdim}.

The dynamical Lévy lattice simulations of Ref.~\cite{aiudi2023critical} determine $z$ directly for the two-dimensional long-range Ising model, without recourse to the effective dimension mapping (see~\Cref{fig:dynamical_exponent_all}). Their estimate is consistent with ours near $\sigma_{\rm mf}$ but a clear discrepancy develops at the Sak boundary: our flow saturates to the short-range value at $\sigma_*(1)=2-\eta_{\rm SR}$, whereas the numerical data follow the free super-diffusive law $z\simeq\sigma$ up to $\sigma=2$. Ref.~\cite{aiudi2023critical} leaves the location of this dynamical crossover open, noting that settling it requires analytical RG input: our calculation supplies such input and places it at the static Sak value. We note that simulations of long-range models in this window are known to be affected by strong finite-size corrections~\cite{brezin2014crossover,angelini2014relations}, and that the dynamical Lévy lattice does not satisfy detailed balance, its validation against known critical exponents having been carried out in equilibrium; its dynamical critical behavior therefore lacks an independent benchmark.

Once more, as in \Cref{subsec:1DLRI}, we have incorporated MC results from Refs.~\cite{grassberger1995damage,nightingale2000monte,hasenbusch2020dynamic} for the two- and three-dimensional short-range Ising models, mapped to the long-range case via \eqref{mapping MC a}, \eqref{effective sigma}, and
\begin{align}\label{mapping MC c}
	&\theta_{\text{LR}}^{\text{MC}} = \theta_{\text{SR}}^{\text{MC}} = {\theta'}_{\text{SR}}^{\text{MC}} - \frac{2-\eta_{\text{SR}}-z_{\text{SR}}^{\text{MC}}}{z_{\text{SR}}^{\text{MC}}}\,.
\end{align}
As seen in~\Cref{fig:slip_exponent_all} and~\ref{fig:dynamical_exponent_all}, these mappings are in reasonable agreement with our fRG values at $\sigma \approx 1.35$ and $\sigma \approx 1.75$, corresponding to $D=3$ and $D=2$, respectively. We emphasize again that there is the possibility of an improved calculation of $\theta$ and $z$, especially for larger $\sigma$, by choosing more involved truncations and cutoff regulators, see the discussion in~\cite{ihssen2025nonperturbative}.

Compared to the one-dimensional case in \Cref{fig:one-dimension}, our method shows an even more pronounced improvement over the $\epsilon$-expansion of Ref.~\cite{chen2000short}. Perturbative RG, by its very nature, fails to capture the correct critical behavior near the short-range crossover at $\sigma=\sigma_*$, where $z \to z_{\rm SR}$ and $\theta \to \theta_{\rm SR}$. Moreover, due to the non-monotonic dependence on $N$ of the function $(N+2)/(N+8)^2$ entering the second-order $\epsilon$-expansion for $z$, the perturbative approach predicts spurious degeneracies such as $z(N=1)=z(N=10)$ for all $\sigma$, which are absent in our non-perturbative treatment.

Finally, starting from the results for the exponents $\theta$ and $z$, we are able to obtain the exponent $\psi = [\theta'+\beta/(\nu z)]^{-1}$ associated with the crossover from short- to long-time universal behavior, as discussed in~\Cref{sec:model}. In fact, due to the scaling relation $\beta/\nu = [\phi] = (d-\sigma)/2$,
\begin{equation}\label{psi exponent}
	\psi = \left(\theta'+\frac{d-\sigma}{2z}\right)^{-1}\,,
\end{equation}
whose large-$N$ limit is $\psi(d,\sigma) \equiv 2$, having used~\eqref{aging exponent largeN}. 
\begin{figure}
	\centering
	\includegraphics[width=1.0\linewidth]{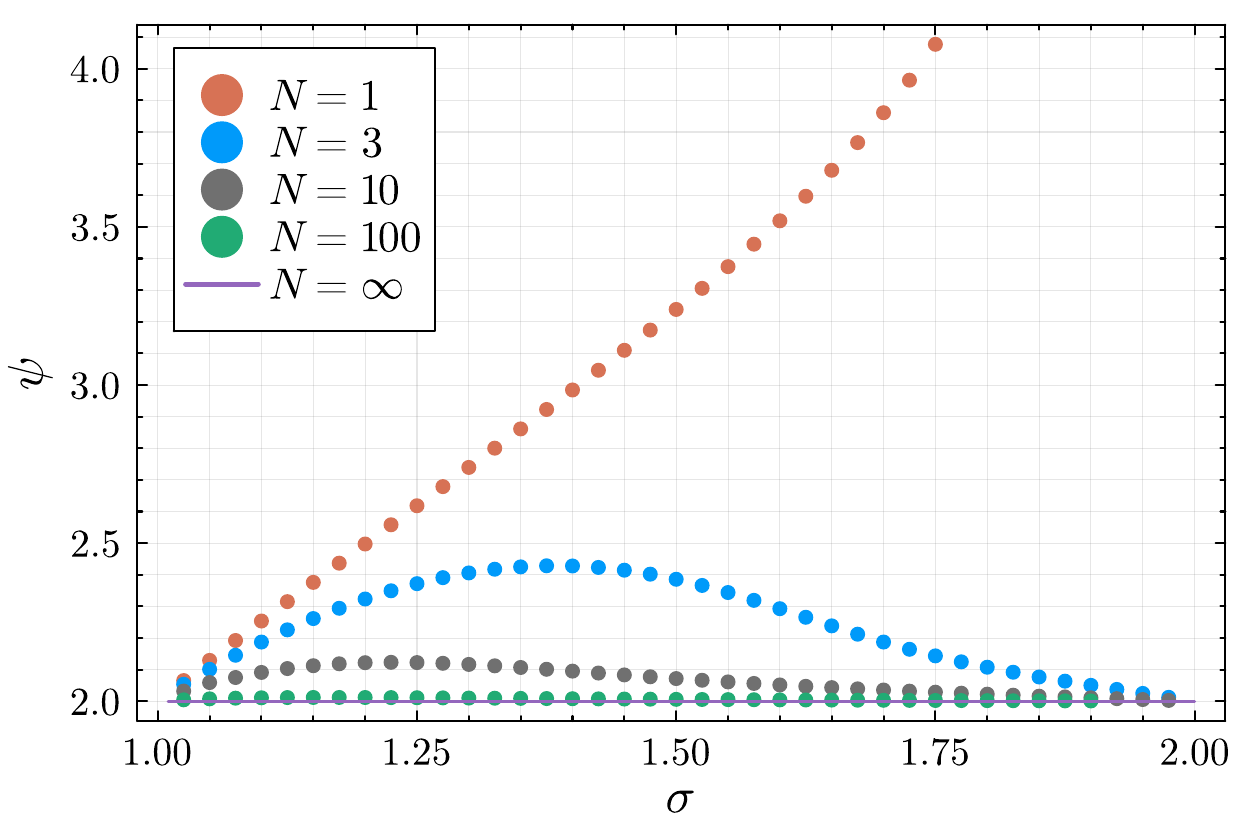}
	\caption{Exponent $\psi$ vs $\sigma$ for the long-range models with $N=1,3,10,100$ in $d=2$. The colored dots represent values obtained from the same fRG data as~\Cref{fig:slip_exponent_all} and~\ref{fig:dynamical_exponent_all}. The large-$N$ limit $\psi = 2$ is plotted as a violet line. 
    \hspace*{\fill}}
	\label{fig:psi_LR}
\end{figure}
In~\Cref{fig:psi_LR} we display the exponent $\psi$ for $d=2$ and several values of $N$. If the initial magnetization $M_0$ is treated as a controllable external parameter, then even small variations in $\psi$ can significantly impact the crossover time $t_{\text{cross}} \propto M_0^{-\psi}$. In particular, for very small $M_0$ one expects a faster crossover to purely relaxational dynamics either when $\sigma$ is close to its mean-field threshold $\sigma_{\rm mf}=d/2$ -- where the performance-rate exponent $\pi_{\rm th}$ is also larger, see~\Cref{fig:performance_rate_exponent} -- or for larger $\sigma$ in the case $N>1$.

\section{Conclusions}\label{sec:conclusions}

In this work, we provided a comprehensive qualitative picture of the dynamics of critical long-range systems using the functional renormalization group, following the methodology of Ref.~\cite{ihssen2025nonperturbative}. Our treatment enabled the derivation of the entire curves of the universal scaling exponents for long-range $O(N)$ models as a function of the symmetry index $N$, the decay exponent $\sigma$, and the dimension $d$.

For the 1D Ising model (\Cref{subsec:1DLRI}), where MC studies are available, our results closely match numerical predictions. For the dynamical critical exponent $z$, MC estimates tend to be slightly overestimated, failing to capture leading-order RG behavior near the mean-field limit $\sigma \simeq 0.5$. In contrast, the fRG curve agrees with perturbative results at small $\sigma$ and deviates towards larger values as $\sigma\to 1$, consistent with the trend set by the MC results. For the aging exponent $\theta'$, MC results appear more reliable: they coincide with perturbative estimates at $\sigma \simeq 0.5$ and remain close to the fRG curve for $\sigma < 0.9$. However, the behavior of critical aging as $\sigma \to 1$ remains unclear, though a non-monotonic trend in scaling indices might be expected from equilibrium studies~\cite{pagni2025one}.

In $d = 2$, to our knowledge, there are no numerical studies of critical aging. However, our results can be compared with the scaling indices obtained via the correspondence with the local model (see~\eqref{effective sigma}), at least for the Ising case. From this perspective, our accuracy remains high up to $\sigma \gtrsim 1.5$, but deviations appear as $\sigma \to \sigma_{*}$. The magnitude of this deviation is consistent with that expected for LPA$'$ in local models~\cite{ihssen2025nonperturbative} in $d = 2$. In the present case, however, it is unclear whether this discrepancy stems from the limitations of the LPA$'$ ansatz or from the approximate nature of the long-range to short-range correspondence~\cite{defenu2015fixed}. In any case, for higher symmetry groups ($N \geq 2$) we expect critical fluctuations to diminish, making the LPA$'$ ansatz increasingly reliable~\cite{codello2015critical}. This trend is clearly demonstrated by the collapse of our results onto the exactly solvable $N = \infty$ case.

Finally, we note that the applicability of the initial-slip exponent is not confined to quenches at criticality. Recent work~\cite{henkel2025physical} shows that the exponent $\lambda$ governing the long-time decay of the autocorrelation function satisfies $\lambda = d-\theta' z$ also for \emph{sub}-critical quenches ($T<T_c$), thereby extending this relation from critical dynamics to the phase-ordering regime -- a result recently confirmed numerically for the 2D Ising universality class in Ref.~\cite{moueddene2026short}. This result establishes a direct link between early-time growth, controlled by the initial-slip exponent $\theta'$, and long-time aging behavior, pointing to a unified scaling description of critical aging and coarsening. Clarifying how this unified picture is modified by long-range interactions constitutes an interesting direction for future work.

The relevance of studying critical aging at finite temperature lies in its potential thermodynamic applications. Indeed, Ref.~\cite{campisi2016power} shows that employing a working medium near criticality can enhance the performance of an Otto cycle. In this context, we demonstrate that long-range interactions provide an additional knob to achieve a thermodynamic advantage: they generally reduce the dynamical exponent $z$, thereby shortening cycle times and improving finite-time performance. As a drawback, the specific heat exponent $\alpha$ decreases under long-range interactions, yet the combined effect still leads to an overall enhancement of the performance-rate exponent $\pi_{\rm th}$ near the mean-field limit (see~\Cref{fig:performance_rate_exponent}). Our findings thus highlight how tuning universal behavior through long-range interactions offers a robust pathway to thermodynamic advantage.

Further thermodynamic applications of long-range many-body systems may arise in quantum thermodynamics~\cite{solfanelli2025universal}, where questions on universal dynamical scaling and critical aging intertwine with fundamental issues of pre-thermalization and equilibration~\cite{chiocchetta2017dynamical}. Addressing these challenges may require extending the LPA$'$ ansatz (cf.~\cite{de2020precision} at equilibrium) to capture the chaotic nature of dynamics beyond one-loop~\cite{sciolla2013quantum}. In this context, it could also be possible to incorporate the effects of (long-range) correlated disorder, whose impact on universality closely resembles that of long-range interactions~\cite{bighin2024universal}.

\begin{acknowledgments}

N. D. is grateful to M. Campisi for useful discussions which inspired this work. V. P. acknowledges G. Piccitto for her useful comments on the role of fluctuations in critical heat engines. This research was funded by the Swiss National Science Foundation (SNSF) grant numbers 200021--207537 and 200021--236722, by the Deutsche Forschungsgemeinschaft (DFG, German Research Foundation) under Germany's Excellence Strategy EXC2181/1-390900948 (the Heidelberg STRUCTURES Excellence Cluster) and the Swiss State Secretariat for Education, Research and Innovation (SERI). Partial support by grant NSF PHY-2309135 to the Kavli Institute for
Theoretical Physics (KITP) is also acknowledged.

\end{acknowledgments}

\appendix
\begingroup
\allowdisplaybreaks

\section{Dynamical scaling in the large-$N$ limit}\label{app:largeN}

In this Appendix we summarize how the decoupling~\eqref{eq:large N decoupling} allows us to obtain an analytic expression for the scaling form of correlation functions and, as a consequence, for the dynamical exponents $z$ and $\theta$. First, the Gaussian correlators are provided. After that, we outline the self-consistent procedure based on~\eqref{time-dependent coupling}.

\subsection{Response and correlation functions}

In the quadratic theory describing the large-$N$ limit it is found (cf.~\cite{janssen1989new}) that, as a generalization of~\eqref{gaussian correlators G},
\begin{equation}\label{response function largeN}
	G^R(\boldsymbol{q},t,t') = \vartheta(t-t') e^{-\int_{t'}^{t} [q^\sigma + \tau_C(u)] \, du}\,,
\end{equation}
where $\tau_0^{-1}=0$ has been set. The bare dispersion relation $q^\sigma + \tau$ has been replaced by $q^\sigma + \tau_C(t)$. On the other hand,
\begin{equation}\label{correlator largeN}
	G^C(\boldsymbol{q},t,t') = 2\int_{t_0}^\infty G^R(\boldsymbol{q},t,u)G^R(\boldsymbol{q},t',u) \, du\,.
\end{equation}
In the infinite-time limit $C(\boldsymbol{q},\infty) = (q^\sigma + \tau_C(\infty))^{-1}$, with $\tau_C(\infty) = \xi^{-2}$, is the Gaussian correlation function in equilibrium. 

\subsection{Self-consistent calculation}

Let us start with the equal-time correlator, whose scaling behavior at criticality ($\tau_C \to 0$) is taken to be the same as that of~\eqref{gaussian correlators C} for $t'=t$, i.e.
\begin{equation}\label{scaling correlator}
	C(\boldsymbol{q},t) = \frac{1}{q^\sigma} F(2 q^\sigma t)\,, \quad F(s)=\begin{cases}
		0 \quad \text{if $s=0$}
		\\
		1 \quad \text{if $s=\infty$}
	\end{cases}\,.
\end{equation}
On the other hand, since $\tau$ and $q^\sigma$ have the same dimensions, we are allowed to make the ansatz
\begin{equation}\label{scaling ansatz tauC}
	\tau_C(t) = \frac{\gamma}{2t}\,,
\end{equation}
provided that $\gamma$ is a dimensionless constant, and in particular time-independent. Now, the crucial point is that the time-dependent coupling and the equal-time correlator have to obey the self-consistency equation~\eqref{time-dependent coupling}, which can be rewritten as
\begin{equation}\label{self consistency}
	\tau_C(t) = \tau_C(\infty) + \frac{g}{6} \int_{\boldsymbol{q}} [C(\boldsymbol{q},t)-C(\boldsymbol{q},\infty)]\,.
\end{equation}
As shown below, this implies that the only consistent solution for $\gamma$ is $\gamma = d/\sigma-2$. Furthermore, we are going to find that the aging exponent $\theta$ is related to $\gamma$.

In fact, inserting~\eqref{scaling ansatz tauC} into~\eqref{correlator largeN} yields a result compatible with~\eqref{scaling correlator}, and in particular allows to determine the scaling function:
\begin{equation}\label{scaling function largeN}
	F(s) = s \int_0^1 dy \, (1-y)^{\gamma} e^{-s y}\,, \quad (\gamma>-1)\,.
\end{equation}
We are now ready to evaluate~\eqref{self consistency} at criticality ($\tau_C(\infty)=0$), which gives the following condition for $\gamma$:
\begin{align}\label{eq:condition_gamma}
	\gamma &= 2t \, \frac{g_*}{6} \frac{\Omega_d}{(2\pi)^d} \int_0^\Lambda dq \, q^{d-1-\sigma} [F(2 q^\sigma t)-1]
	\nonumber\\
	&= \frac{g_*}{6\sigma} \frac{\Omega_d}{(2\pi)^d} (2t)^{\frac{\epsilon}{\sigma}} \int_0^{2 t \Lambda^\sigma} ds \, s^{-\frac{\epsilon}{\sigma}} [F(s)-1]\,,
\end{align}
where $\Lambda$ is an ultraviolet cutoff for momentum integrals, $\Omega_d = 2\pi^{d/2} / \Gamma(d/2)$, $\epsilon = 2\sigma-d > 0$, and $g_*$ is the fixed-point value of the quartic coupling $g$. Next, we split the integral into two parts:
\begin{equation}
	\int_0^{\infty} ds \, s^{-\frac{\epsilon}{\sigma}} [F(s)-1] - \int_{2 t \Lambda^\sigma}^{\infty} ds \, s^{-\frac{\epsilon}{\sigma}} [F(s)-1]\,.
\end{equation}
The first integral must be zero, because otherwise its prefactor in~\eqref{eq:condition_gamma} would give a $t$-dependent contribution $\propto t^{\epsilon/\sigma}$ to $\gamma$, which is ruled out by $\gamma$ being dimensionless:
\begin{equation}
	0 = \int_0^{\infty} ds \, s^{-\frac{\epsilon}{\sigma}} [F(s)-1] \propto \Gamma(\gamma+\epsilon/\sigma)^{-1}\,,
\end{equation}
having integrated by parts and noticed that the boundary terms vanish. 
Thus, we have to pick one of the poles of the Gamma function. The only solution allowed by the condition $\gamma>-1$ in~\eqref{scaling function largeN} is  
\begin{equation}\label{value of alpha}
	\gamma = -\epsilon/\sigma = d/\sigma-2\,.
\end{equation}
Evaluating the second integral in the regime $t \gg (2\Lambda^{\sigma})^{-1}$ (discarding non-universal details at microscopically short times) would also enable the determination of $g_*$~\cite{janssen1989new}.

Using~\eqref{scaling ansatz tauC} and~\eqref{value of alpha}, the critical form of the response function~\eqref{response function largeN} is then 
\begin{equation}
	G^R(\boldsymbol{q},t,t') = \vartheta(t-t') (t/t')^{\theta} e^{-q^\sigma (t-t')}\,,
\end{equation}
with $\theta = -\gamma/2 = 1-d/(2\sigma)$.

\section{Functional derivatives in systems with boundaries}\label{app:variations with boundary}

Let us consider a system on the manifold $\mathcal{M}$, whose boundary is $\mathcal{B} = \partial \mathcal{M}$. In our case the manifold is flat: $\mathcal{M} \subset \mathbb{R}^D$, for some integer $D$. Given a functional
\begin{equation}
	J[\phi] = \int_{\mathcal{M}} dV \, f(\phi,\nabla\phi) + \int_{\mathcal{B}} dA \, g(\phi)\,,
\end{equation}
upon the variation of the function from $\phi$ to $\phi+\delta \phi$
\begin{align}
	&\delta_\phi J = \int_{\mathcal{M}} dV \, \left[ \frac{\partial f}{\partial \phi} - \nabla \cdot \left(\frac{\partial f}{\partial (\nabla \phi)}\right) \right] \delta \phi
	\nonumber\\
	& + \int_{\mathcal{B}} dA \, \left[ \frac{\partial f}{\partial (\nabla \phi)} \cdot \boldsymbol{n} + \frac{\partial g}{\partial \phi} \right]_{\phi = \phi_{\mathcal{B}}} \delta \phi_{\mathcal{B}}\,,
\end{align}
where the function on the boundary is denoted by $\phi_{\mathcal{B}}$ and $\boldsymbol{n}$ is the unit vector normal to the boundary. 
As hinted by the notation, we identify $\phi$ with the order-parameter field $\phi(t,\boldsymbol{x})$, and for simplicity we neglect its spatial dependence. Then, the manifold $\mathcal{M}$ is $[t_0,\infty) \ni t$, the temporal `bulk', while the boundary is just $\mathcal{B} = \{t_0\}$. Denoting $\partial_t \phi \equiv \dot{\phi}$, one has $f = f(\phi,\dot{\phi})$ and
\begin{align}\label{variation time boundary}
	&\delta_\phi J = \int_{t_0}^\infty dt \, \left[ \frac{\partial f}{\partial \phi} - \partial_t \left(\frac{\partial f}{\partial \dot{\phi} }\right) \right] \delta \phi
	\nonumber\\
	& + \left[ - \frac{\partial f}{\partial \dot{\phi}} + \frac{\partial g}{\partial \phi} \right]_{\phi = \phi_0} \delta \phi_0\,.
\end{align}

\subsection{Variations of the effective average action}\label{subsec:variations_EAA}

Clearly, we want to compute the variations of the ansatz~\eqref{EAA ansatz}. Hereafter, let us write $\int_t \equiv \int_{t_0}^\infty$ as a shorthand notation for time integrals. Then
\begin{align}
	& \Gamma[\boldsymbol{\phi},	\tilde{\boldsymbol{\phi}}] = \int_{t} \tilde{\phi}_i (t) \left[Z \dot{\phi}_i(t) + K q^{\sigma} \phi_i(t) - \Omega \tilde{\phi}_i (t) \right.
	\nonumber\\
	&
	\left. + V^{(i)}(\boldsymbol{\phi}(t)) \right] + \left(Z_{0} \tilde{\phi}_{0,i} {\phi}_{0,i} -\frac{Z_{0}^2}{2 \tau_0} \tilde{\phi}_{0,i}^2 \right)\,,
\end{align}
meaning that
\begin{align}
	f(\Phi,\dot{\Phi}) &= \tilde{\phi}_i \left[Z \dot{\phi}_i + K q^{\sigma} \phi_i - \Omega \tilde{\phi}_i + V^{(i)}(\boldsymbol{\phi}) \right]\,,
	\\
	g(\Phi) &= Z_{0} \tilde{\phi}_{i} {\phi}_{i} -\frac{Z_{0}^2}{2 \tau_0} \tilde{\phi}_{i}^2\,,
\end{align}
where concretely one can read $\Phi = \phi_j$ or $\Phi = \tilde{\phi}_j$ according to which field we are choosing when performing the variation. Using~\eqref{variation time boundary}, one obtains the following results:
\begin{subequations}
\begin{align}
	&\delta_{\phi_i} \Gamma = \int_t \left[ \tilde{\phi}_i K q^{\sigma} + \tilde{\phi}_j V^{(ij)}(\boldsymbol{\phi}) - Z \dot{\tilde{\phi}}_i \right] \delta \phi_i
	\nonumber\\
	& + (Z_0-Z) \tilde{\phi}_{0,i} \delta \phi_{0,i}\,,
	\\
	&\delta_{\tilde{\phi}_i} \Gamma = \int_t \left\{ \left[Z \dot{\phi}_i + K q^{\sigma} \phi_i + V^{(i)}(\boldsymbol{\phi}) \right] - 2 \Omega \tilde{\phi}_i \right\} \delta \tilde{\phi}_i
	\nonumber\\
	& + \left(Z_0 \phi_{0,i} - \frac{Z_0^2}{\tau_0} \tilde{\phi}_{0,i}\right) \delta\tilde{\phi}_{0,i}\,,
\end{align}
\end{subequations}
and
\begin{subequations}\label{hessian}
	\begin{align}
		&\delta^2_{\tilde{\phi}_i \tilde{\phi}_j} \Gamma = \delta_{ij} \int_t (- 2\Omega) \delta\tilde{\phi}_i \delta\tilde{\phi}_j - \delta_{ij} \frac{Z_0^2}{\tau_0} \delta\tilde{\phi}_{0,i} \delta{\tilde{\phi}}_{0,j}\,,
		\\
		& \delta^2_{\phi_i \tilde{\phi}_j} \Gamma = \int_t \left\{\left[ K q^\sigma \delta_{ij} + V^{(ij)}(\boldsymbol{\phi}) \right] \delta \phi_i + \delta_{ij} Z \partial_t \delta\phi_i \right\} \delta \tilde{\phi}_j
		\nonumber\\
		& + \delta_{ij} Z_0 \delta\phi_{0,i} \delta \tilde{\phi}_{0,j}\,, \label{del phi tildephi Gamma}
		\\
		&\delta^2_{\tilde{\phi}_i \phi_j} \Gamma = \int_t \left\{ \left[ Kq^\sigma \delta_{ij} + V^{(ij)}(\boldsymbol{\phi}) \right] \delta\tilde{\phi}_i - \delta_{ij} Z \partial_t \delta\tilde{\phi}_i \right\} \delta\phi_j
		\nonumber\\
		& + \delta_{ij} {(Z_0-Z)} \delta\tilde{\phi}_{0,i} \delta{\phi}_{0,j}\,.
	\end{align}
\end{subequations}
The relations~\eqref{hessian}, together with the obvious $\delta^2_{{\phi}_i {\phi}_j} \Gamma$ involving only derivatives of the effective potential, form the Hessian matrix $\Gamma^{(2)}$.

\subsection{Variations of the Wetterich equation}\label{subsec:variations_wetterich}

Let us now consider the right-hand side of~\eqref{wetterich},
\begin{align}
	\frac{\partial_\kappa R}{2} \int_{t} \tr \left[\mathds{G}(t,t) \hat{\tau}_1 \right] \equiv \frac{\partial_\kappa R}{2} \mathcal{J}[\boldsymbol{\phi},	\tilde{\boldsymbol{\phi}}]\,,
\end{align}
where the trace is over the $2\times2$ matrix structure of the response field framework (and possibly over $\mathrm{O}(N)$ indices if $N>1$). Ignoring the prefactor, we make the identifications $f(\Phi) = \tr \left[\mathds{G}(t,t) \hat{\tau}_1 \right]$ (where $f$ is independent of derivatives of the field, as the propagator is so) and $g(\Phi) = 0$. Calculating the first and second variations of the Wetterich equation w.r.t.~$\phi$ and $\tilde{\phi}$ allows us to compare with the results of~\Cref{subsec:variations_EAA} and obtain the flow of the potential~\eqref{flow eqn derivative effective potential} and the other renormalization functions. One immediately sees that
\begin{align}\label{del phi tildephi Wetterich}
	&\delta^2_{\phi_i \tilde{\phi}_j} \mathcal{J}[\boldsymbol{\phi}, \tilde{\boldsymbol{\phi}}] = \int_t \tr\left\{ \delta^2_{\phi_i \tilde{\phi}_j} \mathds{G}(t,t) \hat{\tau}_1 \right\}\,,
\end{align}
and similarly for the other variations, where, due to~\eqref{full propagator},
\begin{align}
	& \delta_{\Phi} \mathds{G}(t_1,t_2) = - \left[\mathds{G} \left( \delta_{\Phi} \Gamma^{(2)} \right) \mathds{G}\right](t_1,t_2)
	\nonumber\\
	&= - \int_{t'} \mathds{G}(t_1,t') \left( \delta_{\Phi} \Gamma^{(2)} \right) \mathds{G}(t',t_2) \,,
\end{align}
since the genuine time-dependence is contained in the propagator matrix, as motivated below. To compare the right- and left-hand side of~\eqref{wetterich} one needs to rewrite $\delta_{\Phi} \Gamma^{(2)} = \int_t \frac{\delta \Gamma^{(2)}}{\delta \Phi} \delta \Phi$, where $\frac{\delta \Gamma^{(2)}}{\delta \Phi}$ is the functional derivative, and notice that the variation $\delta \Phi$ is arbitrary.

\section{Inversion of the Hessian}\label{app:invert hessian}

As discussed in~\Cref{subsec:propagators_projections}, on the stationary background $\boldsymbol{\Phi}_u$ the regulated Hessian decomposes into one longitudinal and $N-1$ identical transverse sectors, with renormalized dispersions given in~\eqref{renormalized dispersions main}. Here, we provide its explicit inversion, first in the time-translation-invariant limit and then in the presence of the initial-time boundary.

In the limit $t_0\to-\infty$, time-translation invariance is restored and the Hessian can be transformed to frequency space. Its $2\times2$ block in correlation-response field space, for each sector $i\in\{L,T\}$, is
\begin{equation} 
    (\mathds{G}_k^{-1})_i = \begin{pmatrix} 0 & P_k^i(\omega,q^\sigma) \\ P_k^i(-\omega,q^\sigma) & -2\Omega_k \end{pmatrix}, 
\end{equation}
where $P_k^i(\omega,q^\sigma) \equiv Z_k i\omega+\omega_{i,k}(q^\sigma)$. Now, the inversion can be performed independently in the longitudinal and transverse sectors, yielding
\begin{equation}\label{TTI propagator} 
    \mathds{G}_{k,i}(\omega,\boldsymbol{q}) = \begin{pmatrix} \frac{2\Omega_k} {P_k^i(\omega,q^\sigma)P_k^i(-\omega,q^\sigma)} & \frac{1}{P_k^i(-\omega,q^\sigma)} \\[0.3cm] \frac{1}{P_k^i(\omega,q^\sigma)} & 0 \end{pmatrix}. 
\end{equation}
An inverse Fourier transform gives the time-translation-invariant correlation and response functions reported in~\eqref{equilibrium propagator matrix}--\eqref{equilibrium correlators}.

At finite $t_0$, time-translation invariance is broken and the Hessian cannot be inverted by a Fourier transform in time. To retain the boundary contribution, we decompose the inverse propagator into a field-independent and a field-dependent part,
\begin{equation}\label{field splitting inverse propagator} 
    \mathds{G}^{-1}(x,x';\boldsymbol{\Phi}) = \mathds{G}_0^{-1}(x,x') + \mathds{V}(x,x';\boldsymbol{\Phi}), 
\end{equation} where $\mathds{V}(x,x';\boldsymbol{\Phi}) = \delta(x-x') \vartheta(t-t_0) \mathds{V}(x;\boldsymbol{\Phi})$ and using the notation $x=(t,\boldsymbol{x})$.
\begin{widetext}
The field-dependent contribution reads 
\begin{equation}
	\mathds{V}_{ij}(x;\boldsymbol{\Phi}) = \begin{pmatrix}
		\tilde{\phi}_l V_k^{(ijl)}(\boldsymbol{\phi}) & \delta_{ij} [V_k^{(ii)}(\boldsymbol{\phi}) - V_k^{(ii)}(\boldsymbol{\phi}_u)]
		\\
		\delta_{ij} [V_k^{(ii)}(\boldsymbol{\phi}) - V_k^{(ii)}(\boldsymbol{\phi}_u)] & 0
	\end{pmatrix}\,,
\end{equation}
while the field-independent part is  obtained from the quadratic terms of $\Gamma_k$ (see below). On the stationary background $\boldsymbol{\Phi}_u=(\boldsymbol{\phi}_u,\boldsymbol{0})$, the field-dependent contribution $\mathds{V}$ vanishes, so that $\mathds{G}^{-1}=\mathds{G}_0^{-1}$. We stress that the splitting in~\eqref{field splitting inverse propagator} is valid also for $t_0\to-\infty$: $\mathds{V}$ vanishes on the stationary background in both temporal regimes.

Then, the only task left is to obtain the explicit form of $\mathds{G}_0$. It proves convenient to treat the boundary and bulk fields as separate components of enlarged field vectors. For each sector $i\in\{L,T\}$ and momentum $\boldsymbol{q}$, we define 
\begin{equation} 
    \Phi \equiv \begin{pmatrix} \phi_0\\ \phi(t) \end{pmatrix}, \qquad \widetilde{\Phi} \equiv \begin{pmatrix} \tilde{\phi}_0\\ \tilde{\phi}(t) \end{pmatrix}, \qquad t>t_0, 
\end{equation} 
with the matching condition $\phi(t_0^+)=\phi_0$. The sector index on the fields is suppressed in the following. For two generic boundary-bulk vectors of the form $X= (X_0 \; X(t))^T$ we define an inner product 
\begin{equation} 
    \langle X,X'\rangle \equiv X_0X_0' + \int_{t_0}^{\infty}dt\,X(t)X'(t). \label{boundary bulk inner product} 
\end{equation}
We introduce the operators $\mathcal{L}_{i,k}$ and $\mathcal{Q}_k$ through their actions on $\Phi$ and $\tilde \Phi$, respectively: 
\begin{equation}
    \mathcal{L}_{i,k}\Phi \equiv \begin{pmatrix} Z_{0,k}\phi_0\\[1mm] \left[ Z_k\partial_t+\omega_{i,k}(q^\sigma) \right]\phi(t) \end{pmatrix},
    \qquad
    \mathcal{Q}_k\widetilde{\Phi} \equiv \begin{pmatrix} \dfrac{Z_{0,k}^2}{2\tau_0}\tilde{\phi}_0\\[3mm] \Omega_k\tilde{\phi}(t) \end{pmatrix}.
\end{equation} 
The quadratic effective action on the stationary background can then be written as 
\begin{align} 
    \Gamma_{0,k}^{(i)} &= \left\langle \widetilde{\Phi}, \mathcal{L}_{i,k}\Phi \right\rangle - \left\langle \widetilde{\Phi}, \mathcal{Q}_k\widetilde{\Phi} \right\rangle = Z_{0,k}\tilde{\phi}_0\phi_0 -\frac{Z_{0,k}^2}{2\tau_0}\tilde{\phi}_0^2 + \int_{t_0}^{\infty}dt\, \left\{ \tilde{\phi}(t) \left[ Z_k\partial_t+\omega_{i,k}(q^\sigma) \right]\phi(t) -\Omega_k\tilde{\phi}(t)^2 \right\}.
    \label{quadratic action operator form} 
\end{align} 
The adjoint $\mathcal{L}_{i,k}^{\dagger}$, defined by $\langle \widetilde{\Phi}, \mathcal{L}_{i,k}\Phi \rangle = \langle \mathcal{L}_{i,k}^{\dagger}\widetilde{\Phi}, \Phi \rangle$, acts on the response-field vector as 
\begin{equation}
    \mathcal{L}_{i,k}^{\dagger}\widetilde{\Phi} = \begin{pmatrix} Z_{0,k}\tilde{\phi}_0 - Z_k\tilde{\phi}(t_0^+) \\[1mm] \left[ -Z_k\partial_t+\omega_{i,k}(q^\sigma) \right]\tilde{\phi}(t) \end{pmatrix}, \label{adjoint response operator} 
\end{equation}
as it is verified by integrating the time-derivative term by parts. 

We now further collect the order-parameter and response-field vectors into $\Psi \equiv (\Phi \ \widetilde{\Phi})^T$ and extend the inner product according to $\langle \Psi,\Psi' \rangle \equiv \langle \Phi,\Phi' \rangle + \langle \widetilde{\Phi},\widetilde{\Phi}' \rangle$. The quadratic action thus becomes 
\begin{equation} 
    \Gamma_{0,k}^{(i)} = \frac{1}{2} \left\langle \Psi, \mathcal{A}_{i,k}\Psi \right\rangle, \qquad \mathcal{A}_{i,k} \equiv \mathcal{G}_{0,i}^{-1} = \begin{pmatrix} 0 & \mathcal{L}_{i,k}^{\dagger}\\ \mathcal{L}_{i,k} & -2\mathcal{Q}_k \end{pmatrix}. \label{enlarged inverse propagator}
\end{equation} 
The inverse of this block operator is
\begin{equation} 
    \mathcal{G}_{0,i} \equiv \begin{pmatrix} \mathcal{G}_{0,i}^{C} & \mathcal{G}_{0,i}^{R}\\ \mathcal{G}_{0,i}^{A} & 0 \end{pmatrix} = \mathcal{A}_{i,k}^{-1} = \begin{pmatrix} 2\mathcal{L}_{i,k}^{-1} \mathcal{Q}_k \mathcal{L}_{i,k}^{-\dagger} & \mathcal{L}_{i,k}^{-1} \\[1mm] \mathcal{L}_{i,k}^{-\dagger} & 0 \end{pmatrix}, 
    \qquad \text{with }\mathcal{L}_{i,k}^{-\dagger} \equiv ( \mathcal{L}_{i,k}^{\dagger} )^{-1} = ( \mathcal{L}_{i,k}^{-1})^{\dagger}.
    \label{enlarged propagator} 
\end{equation}
The enlarged propagator $\mathcal{G}_{0,i}$ contains all boundary-boundary, boundary-bulk, and bulk-bulk correlation and response functions. The propagator $\mathds{G}_{0,i}(t,t')$ used in the flow equations is its restriction to the bulk sector: \begin{equation} 
    \mathds{G}_{0,ij}(t,t') = \delta_{ij} \begin{pmatrix}
		G^C_{0,i}(t,t') & G^R_{0,i}(t,t') \\
		G^R_{0,i}(t',t) & 0
	\end{pmatrix}, \qquad t,t'>t_0. \label{Gaussian propagator matrix} 
\end{equation} 

To determine $\mathcal{L}_{i,k}^{-1}$, we solve the linear problem $\mathcal{L}_{i,k}\Phi = J$, with $J=(j_0 \ j(t))^T$. In components, this gives 
\begin{subequations} 
\begin{align} 
    Z_{0,k}\phi_0&=j_0,\\ \left[ Z_k\partial_t+\omega_{i,k}(q^\sigma) \right]\phi(t)&=j(t), \qquad t>t_0, 
\end{align}
\end{subequations} 
together with the matching condition $\phi(t_0^+)=\phi_0$. Its solution is $\phi_0 = j_0/Z_{0,k}$ and 
\begin{align} 
    \phi(t) &= \frac{j_0}{Z_{0,k}} e^{-\omega_{i,k}(q^\sigma)(t-t_0)/Z_k} + \frac{1}{Z_k} \int_{t_0}^{\infty}dt'\, \vartheta(t-t') e^{-\omega_{i,k}(q^\sigma)(t-t')/Z_k} j(t'). 
\end{align}
Therefore, 
\begin{equation} 
    \mathcal{L}_{i,k}^{-1} = \begin{pmatrix} \dfrac{1}{Z_{0,k}} & 0 \\[3mm] \dfrac{e^{-\omega_{i,k}(q^\sigma)(t-t_0)/Z_k}}{Z_{0,k}} & \dfrac{\vartheta(t-t') e^{-\omega_{i,k}(q^\sigma)(t-t')/Z_k}}{Z_k} \end{pmatrix}, \qquad \mathcal{L}_{i,k}^{-\dagger} = \begin{pmatrix} \dfrac{1}{Z_{0,k}} & \dfrac{e^{-\omega_{i,k}(q^\sigma)(t'-t_0)/Z_k}}{Z_{0,k}} \\[3mm] 0 & \dfrac{\vartheta(t'-t) e^{-\omega_{i,k}(q^\sigma)(t'-t)/Z_k}}{Z_k} \end{pmatrix}. \label{inverse finite time response operator} 
\end{equation} 
The bulk-bulk terms of~\eqref{inverse finite time response operator} immediately give, through~\eqref{enlarged propagator}, 
\begin{align} 
    G^R_{0,i}(t,t') &= \frac{\vartheta(t-t')}{Z_k} e^{-\omega_{i,k}(q^\sigma)(t-t')/Z_k}, \qquad G^A_{0,i}(t,t') = \frac{\vartheta(t'-t)}{Z_k} e^{-\omega_{i,k}(q^\sigma)(t'-t)/Z_k} = G^R_{0,i}(t',t). 
\end{align} \label{finite time response functions} 
Finally, a short calculation reveals that the bulk-bulk component of $\mathcal{G}_{0,i}^{C} = 2\mathcal{L}_{i,k}^{-1} \mathcal{Q}_k \mathcal{L}_{i,k}^{-\dagger}$ is given by 
\begin{align}
    G^C_{0,i}(t,t') &= \frac{1}{\tau_0} e^{-\omega_{i,k}(q^\sigma)(t+t'-2t_0)/Z_k} + \frac{2\Omega_k}{Z_k^2} \int_{t_0}^{\min(t,t')}ds\, e^{-\omega_{i,k}(q^\sigma)(t+t'-2s)/Z_k} \\
    &= \frac{\Omega_k} {Z_k\omega_{i,k}(q^\sigma)} e^{-\omega_{i,k}(q^\sigma)|t-t'|/Z_k} + \left[ \frac{1}{\tau_0} - \frac{\Omega_k} {Z_k\omega_{i,k}(q^\sigma)} \right] e^{-\omega_{i,k}(q^\sigma)(t+t'-2t_0)/Z_k}. \label{gaussian correlation function} 
\end{align} 
Compared to the equilibrium correlation and response functions~\eqref{equilibrium correlators}, only the correlator receives a correction that breaks time-translation invariance, while $G^R_{0,i}(t,t') = G^R_{\text{eq},i}(t,t')$. This correction,
\begin{equation} G^C_{\mathrm{neq},i}(t,t') = \frac{\gamma_{i,k}}{\omega_{i,k}(q^\sigma)} e^{-\omega_{i,k}(q^\sigma)(t+t'-2t_0)/Z_k}, \qquad \gamma_{i,k} \equiv \frac{\omega_{i,k}(q^\sigma)}{\tau_0} - \frac{\Omega_k}{Z_k}, \end{equation}
vanishes in the limit $t_0 \to -\infty$ or $t+t' \to \infty$. Moreover, independent of the values of $Z_k, \Omega_k, Z_{0,k}$, one finds $G^C_{0,i}(t_0,t_0)=\frac{1}{\tau_0}$, as required by the prescribed initial distribution~\eqref{eq:correlators_initial_time}.

\section{Calculation of the flow equations}\label{app:calculation of the flows}

Throughout this Appendix, we work directly in the case $\tau_0 = \infty$ (cf.~\Cref{sec:model}). Thus, the coefficient $\gamma_{i,k}$ in the non-equilibrium part of~\eqref{gaussian correlation function} becomes independent of the $N$-component: $\gamma_{i,k} = \gamma_k$. We illustrate here only the calculation of the flows of $Z_{0,k}$ and $\Omega_k$, but those of the other quantities are obtained in a very similar manner.

\subsection{Flow of $Z_0$}
Let us consider the RG-time derivatives of~\eqref{del phi tildephi Gamma} and~\eqref{del phi tildephi Wetterich} on the temporal boundary. Comparing them yields
\begin{align}\label{eq_test}
	\delta_{ij} \partial_\kappa Z_0 \, \delta\phi_{0,i} \delta \tilde{\phi}_{0,j} &= \frac{\partial_\kappa R}{2}  \int_t \tr\left\{ \delta^2_{\phi_i \tilde{\phi}_j} \mathds{G}(t,t) \hat{\tau}_1 \right\} \equiv \frac{\partial_\kappa R}{2} \mathcal{I}_{ij}\,.
\end{align}
We immediately see that in order to find $ \delta\phi_{0,i}$ and $\delta \tilde{\phi}_{0,i}$ on the right-hand side of the previous equation we need to use the procedure described in~\Cref{app:trick_short_time}. Let us split the integral as $\mathcal{I}_{ij} = \mathcal{I}_{ij}^{\text{a}} + \mathcal{I}_{ij}^{\text{b}}$, with
\begin{subequations}
\begin{align}
	&\mathcal{I}_{ij}^{\text{a}} = \sum_{a,b=1}^N \int_{t,t',t''} \tr \left\{
	\left[
	\mathds{G}_a(t,t'') (\delta_{\tilde{\phi}_j}\Gamma^{(2)})_{ab} \mathds{G}_b(t'',t') (\delta_{\phi_i}\Gamma^{(2)})_{ba} \mathds{G}_a(t',t)
	\right. \right.
	\nonumber\\
	&
	\left.\left.
	+ \mathds{G}_a(t,t') (\delta_{\phi_i}\Gamma^{(2)})_{ab} \mathds{G}_b(t',t'') (\delta_{\tilde{\phi}_j}\Gamma^{(2)})_{ba} \mathds{G}_a(t'',t)
	\right] \hat{\tau}_1
	\right\}
	\\
	& \mathcal{I}_{ij}^{\text{b}} = - \sum_{a=1}^N \int_{t,t'}  \tr\left\{
	\mathds{G}_a(t,t') (\delta^2_{\phi_i \tilde{\phi}_j} \Gamma^{(2)})_{aa} \mathds{G}_a(t',t) \hat{\tau}_1
	\right\} \,.
\end{align}
\end{subequations}
For practical convenience we compute the integrals over $t$ and $t''$ first, while leaving the $t'$-integral for later. We find that the time-dependence enters only through the combination $(t'-t_0)$. In particular,
\begin{align}
	\mathcal{I}_{ii}^{\text{b}} = \sum_{a=1}^N \frac{V^{(aaii)}(\boldsymbol{\phi})}{\omega_a^2 Z_k} \int_{t'} \left(e^{-2\omega_a (t'-t_0)/Z_k}
	[\Omega_k - 2 \gamma \omega_a (t'-t_0)]-\Omega_k\right) \delta \phi_i \delta\tilde{\phi}_i\,,
\end{align}
and a much longer expression for $\mathcal{I}_{ii}^{\rm a}$, which we do not display here, but it is again an integral over $t'$ where the time-dependent part of the integrand is proportional to $e^{-3(\omega_a+\omega_b)(t'-t_0)/Z}$. We can now employ the short-time trick described in~\Cref{app:trick_short_time} to obtain the variation at the time-boundary $t'=t_0$ and compare the result with the left-hand side. Using~\eqref{trick taylor expansion}, we obtain
\begin{subequations}
\begin{align}
	&\mathcal{I}_{ii}^{\text{b}}(t_0) = \sum_{a=1}^N \frac{V^{(aaii)}(\boldsymbol{\phi})}{2 \omega_a^3}  
	(\Omega_k- Z_k \gamma) \delta \phi_{0,i} \delta\tilde{\phi}_{0,i} \,,
	\\
	&\mathcal{I}_{ii}^{\text{a}}(t_0) = - \sum_{a,b=1}^N V^{(abi)}(\boldsymbol{\phi})^2  \frac{\omega _b^2 \Omega_k \left(\omega _a+\omega _b\right) -  Z_k \gamma \left(2 \omega _a \omega _b^2+\omega _a^3+\omega
		_b^3\right)}{\omega _a^3 \omega _b^2
		\left(\omega _a+\omega _b\right)^2} \delta \phi_{0,i} \delta\tilde{\phi}_{0,i} \,,
\end{align}
\end{subequations}
once we evaluate on a constant field configuration (in particular with $\tilde{\boldsymbol\phi} = 0$). Finally, if $i=m=1$
\begin{align}
	&\partial_\kappa Z_0 = \int_{\boldsymbol{q}} \frac{\partial_\kappa R}{2}  \left[ - V_k^{(111)}(\boldsymbol{\phi})^2 \frac{\Omega_k - 2 Z_k \gamma}{2 \omega_L^4} - (N-1) V_k^{(1gg)}(\boldsymbol{\phi})^2 \frac{\Omega_k -  2 Z_k \gamma}{2 \omega_T^4} \right.
	\nonumber\\
	& \left. 
	+  \frac{\Omega_k- Z_k \gamma}{2} \left\{ \frac{V_k^{(1111)}(\boldsymbol{\phi})}{\omega_L^3} + (N-1) \frac{V_k^{(11gg)}(\boldsymbol{\phi})}{\omega_{T}^3} \right\}
	\right]\,,
\end{align}
or if $i=g\neq 1$,
\begin{align}
	&\partial_\kappa Z_0 = \int_{\boldsymbol{q}} \frac{\partial_\kappa R}{2}  \left[ - V_k^{(1gg)}(\boldsymbol{\phi})^2 \frac{\Omega_k (\omega_L^4+\omega_L^3 \omega_T + \omega_L \omega_T^3 + \omega_T^4) - \gamma Z_k (\omega_L^4+3\omega_L^3 \omega_T + 3\omega_L \omega_T^3 + \omega_T^4) }{(\omega_L+\omega_T)^2 \omega_L^3 \omega_T^3}
	\right.
	\nonumber\\
	& \left. + \frac{\Omega_k- Z_k \gamma}{2}  \left\{ \frac{V_k^{(1111)}(\boldsymbol{\phi})}{\omega_L^3}
	+ (N+1) \frac{V_k^{(gghh)}(\boldsymbol{\phi})}{\omega_T^3} \right\} \right] \quad  \mathrm{for} \quad h \neq g, h \neq 1\,.
\end{align}

Following the expansion introduced in Ref.~\cite{ihssen2025nonperturbative}, at the leading iteration the propagators appearing in the loop integrals generated by the right-hand side of the Wetterich equation are evaluated by setting $\Omega_k=Z_k=Z_{0,k}$. The flows of these three quantities are nevertheless obtained from different projections of the Wetterich equation. In particular, the boundary projection gives $\eta_{Z_0}\neq\eta_Z$. A higher iteration of the scheme in Ref.~\cite{ihssen2025nonperturbative} would use the resulting distinct values of the renormalization functions to update the propagators inside the loop integrals. 

Here we work at the level of the leading iteration and find
\begin{subequations}\label{etaZ0 complete}
\begin{align}
	&\eta_{Z_0}^{(m)} = - \int_{\boldsymbol{q}} \frac{\partial_\kappa R}{2}  \left\{
	\frac{3 U_k''(\rho) + 12\rho U_k^{(3)}(\rho) + 4\rho^2 U_k^{(4)}(\rho)}{\omega_L^3} + (N-1) \frac{U_k''(\rho) + 2\rho U_k^{(3)}(\rho)}{\omega_{T}^3}
	\right.
	\nonumber\\
	& 
	\left.
	- 3 \rho \left[ \frac{[3 U_k''(\rho) + 2 \rho U_k'''(\rho)]^2}{\omega_L^4} + (N-1) \frac{U_k''(\rho)^2}{\omega_T^4} \right]
	\right\},
    \label{etaZ0 complete massive}
	\\
	&\eta_{Z_0} ^{(g)}= - \int_{\boldsymbol{q}} \frac{\partial_\kappa R}{2}  \left\{
	\frac{U_k''(\rho) + 2\rho U_k^{(3)}(\rho)}{\omega_L^3}
	+ (N+1) \frac{U_k''(\rho)}{\omega_T^3} - 4 \rho U_k''(\rho)^2 \frac{(\omega_L^4+2\omega_L^3 \omega_T + 2\omega_L \omega_T^3 + \omega_T^4) }{(\omega_L+\omega_T)^2 \omega_L^3 \omega_T^3} \right\}\,,
    \label{etaZ0 complete goldstone}
\end{align}
\end{subequations}
where the integration over momenta has finally been reintroduced, as both $R_k$ and $\omega_{i,k}$ depend on the modulus of $\boldsymbol{q}$.

\subsection{Flow of $\Omega$}

Similarly, if we are interested in the flow of $\Omega_k$, we find
\begin{align}
	& - 2 \int_t  \partial_\kappa \Omega \, \delta\tilde{\phi}_i \delta\tilde{\phi}_i = \frac{\partial_\kappa R}{2}  \int_t \tr\left\{ \delta^2_{\tilde{\phi}_i \tilde{\phi}_i} \mathds{G}(t,t) \hat{\tau}_1 \right\}\,,
\end{align}
where the trace on the right-hand side is given by
\begin{align}
	&  \sum_{a,b=1}^N \int_{t',t''} \tr \left\{
	\left[
	\mathds{G}_a(t,t'') (\delta_{\tilde{\phi}_j}\Gamma^{(2)})_{ab} \mathds{G}_b(t'',t') (\delta_{\tilde{\phi}_i}\Gamma^{(2)})_{ba} \mathds{G}_a(t',t)
	+ \mathds{G}_a(t,t') (\delta_{\tilde{\phi}_i}\Gamma^{(2)})_{ab} \mathds{G}_b(t',t'') (\delta_{\tilde{\phi}_j}\Gamma^{(2)})_{ba} \mathds{G}_a(t'',t)
	\right] \hat{\tau}_1
	\right\}\,,
\end{align}
and the remaining trace is over the $2\times2$ matrix structure only. We can perform the latter and the two time integrals to obtain
\begin{equation}
	\sum_{a,b=1}^N \left\{\frac{4 \Omega_k^2}{Z_k} \frac{[2-e^{-(\omega_a+\omega_b)(t-t_0)/Z_k}] \, \omega_a + \omega_b}{\omega_a^2 \omega_b (\omega_a+\omega_b)^2} + \gamma_k \mathcal{E}_1 + \gamma_k^2 \mathcal{E}_2 \right\} V^{(abi)}(\boldsymbol{\phi})^2 \delta\tilde{\phi}_i \delta\tilde{\phi}_i\,,
\end{equation}
where $\mathcal{E}_1$ and $\mathcal{E}_2$ denote terms that are exponentially suppressed in the variable $(t-t_0)$. More precisely, they are sums of several terms proportional to $e^{-\alpha (t-t_0)}$ with some $\alpha>0$. We are going to ignore terms of this kind. On the other hand, the time-dependence enters only through the latter, so we are effectively approximating $\Omega_k$ (and indeed the other renormalization functions as well as the potential) as time-independent quantities. As a byproduct of this observation, it is expected that deviations from the fluctuation-dissipation theorem manifesting as differences in the flow of $Z_k$ and $\Omega_k$ decay exponentially fast in time. We finally get
\begin{align}
	& - \frac{\partial_\kappa \Omega_k}{\Omega_k} = \frac{\Omega_k}{Z_k} \sum_{a,b=1}^N V^{(abi)}(\boldsymbol{\phi})^2 \int_{\boldsymbol{q}} \partial_\kappa R \frac{2 \omega_a + \omega_b}{\omega_a^2 \omega_b (\omega_a+\omega_b)^2}\,,
\end{align}
which, upon choosing $i = g\neq 1$, leads to
\begin{align}\label{goldstone etaZ}
	& \partial_\kappa \Omega_{k} = - \Omega_k {V_k^{(1gg)}}^2 \int_{\boldsymbol{q}} \partial_\kappa R_k \frac{\omega_L^2 + 4 \omega_L \omega_T + \omega_T^2}{\omega_L^2 \omega_T^2 (\omega_L+\omega_T)^2}\,,
\end{align} 
where $V_k^{(111)} = \sqrt{2\rho} [3 U_k''(\rho) + 2 \rho U_k'''(\rho)]$ and $V_k^{(1gg)} = \sqrt{2\rho} U_k''(\rho)$.

\section{Threshold functions}\label{subsec:threshold functions}
For the derivation of flow equations we have defined the dimensionless variable $\tilde{y} \equiv q^\sigma/k^\sigma$ and then used
\begin{align}\label{substitution rule LThresh LR}
	\int_{\boldsymbol{q}} \frac{\partial_\kappa R_k(q^{\sigma})}{[K_k q^{\sigma} + R_k(q^{\sigma}) + K_k k^\sigma w]^{n+1}} &= \frac{4 v_d k^d}{(K_k k^{\sigma})^n} \frac{2}{\sigma} \frac{L_n^{(d,\sigma)}(w)}{(n+\delta_{n,0})}\,,
	\qquad L_n^{(d,\sigma)}(w) \coloneqq \frac{n+\delta_{n,0}}{2} \int_0^\infty d\tilde{y} \frac{ \tilde{y}^{\frac{d}{\sigma}-1} \, s(\tilde{y})}{[p(\tilde{y})+w]^{n+1}}\,.
\end{align}
The functions $L_n^{(d,\sigma)}(\cdot)$ are called threshold functions. We have used $\int_{\boldsymbol{q}} \equiv \int \frac{d^d q}{(2\pi)^d} = \frac{4 v_d k^d}{\sigma} \int_0^\infty d \tilde{y} \, \tilde{y}^{\frac{d}{\sigma}-1}$. Moreover, we have introduced the dimensionless shape function $r(\tilde{y})$, defined by $R_k(q^{\sigma}) \eqqcolon K_k q^{\sigma} r(\tilde{y})$,
and the auxiliary quantities $s(\tilde{y})$ and $p(\tilde{y})$:
\begin{align}
	& \partial_\kappa R_k(q^{\sigma}) \eqqcolon K_k k^{\sigma} s(\tilde{y}) \implies s(\tilde{y}) = -\tilde{y} \left[\eta_K r(\tilde{y}) + \sigma \tilde{y} r'(\tilde{y})\right], \qquad \qquad
	p(\tilde{y}) \coloneqq \tilde{y}(1+r(\tilde{y}))\,.
\end{align}
If we choose the flat cutoff~\eqref{Litim regulator},
\begin{align}
	& r(\tilde{y}) = \frac{(1-\tilde{y}) \, \vartheta(1-\tilde{y})}{\tilde{y}}\,,
	\qquad s(\tilde{y}) = \left[\sigma -\eta_K (1-\tilde{y}) \right]  \vartheta(1-\tilde{y})\,,
	\qquad p(\tilde{y}) = \tilde{y} + (1-\tilde{y}) \, \vartheta(1-\tilde{y}) = 1\,, \ \text{if $\tilde{y} < 1$\,.}
\end{align}
Thus, the corresponding threshold functions are
\begin{align}\label{threshL for Litim LR}
	L_n^{(d,\sigma)}(w) &= \frac{\sigma^2}{2d} \left(1-\frac{\eta_K}{d+\sigma}\right) \frac{n+\delta_{n,0}}{(1+w)^{n+1}}\,,
\end{align}
which reduce to the well-known functions of the theory with momentum dependence $\sim q^2$~\cite{berges2000nonperturbative} in the limit $\sigma = 2$.
\end{widetext}

\section{Extraction of the short-time behavior}\label{app:trick_short_time}

In general, we have to evaluate integrals of the shape
\begin{equation}
	J \coloneqq \int_{t_0}^{\infty} dt \, \chi(\Phi(t)) f(t-t_0) e^{- \varpi (t-t_0)}\,,
\end{equation}
where $\varpi \equiv 2 \omega_k(q)/Z_k$, $\chi$ is a function of time through the fields $\phi, \tilde{\phi}$, and all our cases are covered by
\begin{equation}
	f(t) = 
	1 + a \frac{\varpi t}{2} + b \left(\frac{\varpi t}{2}\right)^2  
	+ c_{\tau_0}(t)\,, \quad \, c_{\tau_0} = O(\tau_0^{-1})\,.
\end{equation}
Note that $c_{\tau_0}(0) = 0$, so that the property $f(0) = 1$ is guaranteed. Moreover, $c_{\tau_0}(t)$ is also at most $O(t^2)$. Omitting the dependence on all parameters but $t$, we define
\begin{equation}
	g(t) \equiv \chi(\Phi(t)) f(t-t_0)\,.
\end{equation}
For any well-behaved function $g(t)$ that admits a Taylor expansion around $t=t_0$ and any $C>0$ satisfies
\begin{equation}\label{trick taylor expansion}
	\int_{t_0}^{\infty} dt \, g(t) e^{-C(t-t_0)} = \sum_{n=0}^{\infty} \frac{1}{C^{n+1}} g^{(n)}(t_0)\,.
\end{equation}
Thus, using the identity~\eqref{trick taylor expansion}, we get
\begin{equation}
	J = \sum_{n=0}^{\infty} \frac{1}{\varpi^{n+1}} \partial_t^n g(t)|_{t=t_0}\,.
\end{equation}
Next, the general Leibniz rule reads
\begin{equation}
	\partial_t^n g(t) = \sum_{k=0}^n \binom{n}{k} \partial_t^{n-k} \chi(\Phi(t)) \, \partial_t^k f(t-t_0)\,,
\end{equation}
for the $n$th derivative of $g(t)$. This implies that for $n\geq 2$ 
\begin{align}
	& \partial_t^n g(t)|_{t=t_0} = 
	\Bigg[ \frac{n(n-1)}{2} \partial_t^{n-2} \chi(\Phi(t)) \, \partial_t^2 f(t-t_0) 
	\nonumber\\
	&  \left.
	+ n \partial_t^{n-1} \chi(\Phi(t)) \, \partial_t f(t-t_0)
	+ \partial_t^{n} \chi(\Phi(t)) \Bigg] \right|_{t=t_0}\,,
\end{align}
because $\partial_t^k f(t-t_0)|_{t=t_0} = 0$ for all $k > 2$. One concludes
\begin{align}\label{result of the trick}
	J &= \chi(\Phi(t_0)) \left[ \frac{1}{\varpi}
	+ \frac{f'(0)}{\varpi^{2}} + \frac{f''(0)}{\varpi^{3}} \right] + \dots
	\nonumber\\
	&= \frac{1
		+ a/2 + b/2}{\varpi} \chi(\Phi(t_0)) + \dots \, .
\end{align}
where the dots denote all terms with at least one time derivative of the fields and are therefore negligible for the flow of $Z_{0,k}$ studied in~\Cref{app:calculation of the flows}. We have also left out the $O(\tau_{0}^{-1})$ terms, which evaluate to zero at the fixed point due to their dimensionality.

\bibliography{ref_lib}

\end{document}